\documentclass[a4paper, onecolumn, accepted=2021-03-29]{quantumarticle}
\pdfoutput=1

\usepackage[utf8]{inputenc}

\usepackage{amsmath}
\usepackage{amssymb}
\usepackage{amsthm}
\usepackage{amsfonts}

\usepackage[numbers,sort&compress]{natbib}
\newcommand{\tableheadingfont}{}

\usepackage{color}

\usepackage[colorlinks]{hyperref}
\usepackage[all]{hypcap}

\usepackage{multirow}
\usepackage{tabularx}

\newcommand{\eq}[1]{\hyperref[eq:#1]{Equation~\ref*{eq:#1}}}
\renewcommand{\sec}[1]{\hyperref[sec:#1]{Section~\ref*{sec:#1}}}
\DeclareRobustCommand{\app}[1]{\hyperref[app:#1]{Appendix~\ref*{app:#1}}}
\newcommand{\fig}[1]{\hyperref[fig:#1]{Figure~\ref*{fig:#1}}}
\newcommand{\tbl}[1]{\hyperref[tbl:#1]{Table~\ref*{tbl:#1}}}

\input{qcircuit}

\newcommand{\lenexp}{{n_e}}
\newcommand{\devoff}{{\delta_{\text{off}}}}
\newcommand{\gexp}{{c_{\text{exp}}}}
\newcommand{\gmul}{{c_{\text{mul}}}}
\newcommand{\gsep}{{c_{\text{sep}}}}
\newcommand{\gpad}{{c_{\text{pad}}}}
\newcommand{\distone}{{d_1}}
\newcommand{\disttwo}{{d_2}}
\newcommand{\productreg}{x}
\newcommand{\workreg}{y}
\newcommand{\gen}{g}

\begin{document}
\title{How to factor 2048 bit RSA integers in 8 hours using 20 million noisy qubits}

\author{Craig Gidney}
\email{craiggidney@google.com}
\affiliation{Google Inc., Santa Barbara, California 93117, USA}
\author{Martin Ekerå}
\affiliation{KTH Royal Institute of Technology, SE-100 44 Stockholm, Sweden \\ Swedish NCSA, Swedish Armed Forces, SE-107 85 Stockholm, Sweden}

\begin{abstract}
We significantly reduce the cost of factoring integers and computing discrete logarithms in finite fields on a quantum computer by combining techniques from Shor 1994, Griffiths-Niu 1996, Zalka 2006, Fowler 2012, Ekerå-Håstad 2017, Ekerå 2017, Ekerå 2018, Gidney-Fowler 2019, Gidney 2019.
We estimate the approximate cost of our construction using plausible physical assumptions for large-scale superconducting qubit platforms:
 a planar grid of qubits with nearest-neighbor connectivity,
 a characteristic physical gate error rate of $10^{-3}$,
 a surface code cycle time of 1 microsecond,
 and a reaction time of 10 microseconds.
We account for factors that are normally ignored such as noise, the need to make repeated attempts, and the spacetime layout of the computation.
When factoring 2048 bit RSA integers, our construction's spacetime volume is a hundredfold less than comparable estimates from earlier works (Van Meter et al. 2009, Jones et al. 2010, Fowler et al. 2012, Gheorghiu et al. 2019).
In the abstract circuit model (which ignores overheads from distillation, routing, and error correction) our construction uses $3 n + 0.002 n \lg n$ logical qubits, $0.3 n^3 + 0.0005 n^3 \lg n$ Toffolis, and $500 n^2 + n^2 \lg n$ measurement depth to factor $n$-bit RSA integers.
We quantify the cryptographic implications of our work, both for RSA and for schemes based on the DLP in finite fields.
\end{abstract}

\maketitle

\section{Introduction}
\label{sec:introduction}
Peter Shor's introduction in 1994 of polynomial time quantum algorithms for factoring integers and computing discrete logarithms \cite{shor1994,Timeline} was a historic milestone that greatly increased interest in quantum computing.
Shor's algorithms were the first quantum algorithms that achieved a superpolynomial speedup over classical algorithms, applied to problems outside the field of quantum mechanics, and had obvious applications.
In particular, Shor's algorithms may be used to break the RSA cryptosystem \cite{rsa} based on the hardness of factoring integers that are the product of two similarly-sized primes (hereafter ``RSA integers"), and cryptosystems based on the discrete logarithm problem (DLP), such as the Diffie-Hellman key agreement protocol \cite{diffie-hellman} and the Digital Signature Algorithm \cite{fips-186-4}.

The most expensive operation performed by Shor's factoring algorithm is a modular exponentiation.
Modern classical computers can perform modular exponentiations on numbers with thousands of bits in under a second.
These two facts may at first glance appear to suggest that factoring a thousand bit number with Shor's algorithm should only take seconds, but unfortunately (or perhaps fortunately), that is not the case.
The modular exponentiation in Shor's algorithm is performed over a superposition of exponents, meaning a quantum computer is required, and quantum hardware is expected to be many orders of magnitude noisier than classical hardware~\cite{schroeder2009dram,Bare13,Kim2014}.
This noise necessitates the use of error correction, which introduces overheads that ultimately make performing reliable arithmetic on a quantum computer many orders of magnitude more expensive than on classical computers \cite{fowler2012surfacecodereview, campbell2018constraintsatisfaction}.

Although Shor's algorithms run in polynomial time, and although there has been significant historical work focusing on reducing the cost of Shor's algorithms and large scale quantum computing architectures \cite{vedral1996arithmetic,beckman1996efficient,cleve2000fast,oskin2002practical,copsey2003toward,bravyi2005distillation,whitney2009fault,van2010distributed,jones2012layered,pavlidis2012fast,fowler2012time,fowler2012surfacecodereview,horsman2012latticesurgery,fowler2018,ekeraa2016modifying} (summarized in \cite{vanquantumcomputerarchitecture}), the constant factors hidden by the asymptotic notation remain substantial.
These constant factors must be overcome, by heavy optimization at all levels, in order to make the algorithms practical.
Current quantum computers are far from being capable of executing Shor's algorithms for cryptographically relevant problem sizes.

\subsection{Our contributions and a summary of our results}

In this work, we combine several novel and existing optimizations to reduce the cost of implementing Shor's algorithms.
The main hurdle to overcome is to implement one or more modular exponentiations efficiently, as these exponentiations dominate the overall cost of Shor's algorithms.

\begin{figure}[p]
    \begin{center}
    \includegraphics[width=1.0\textwidth]{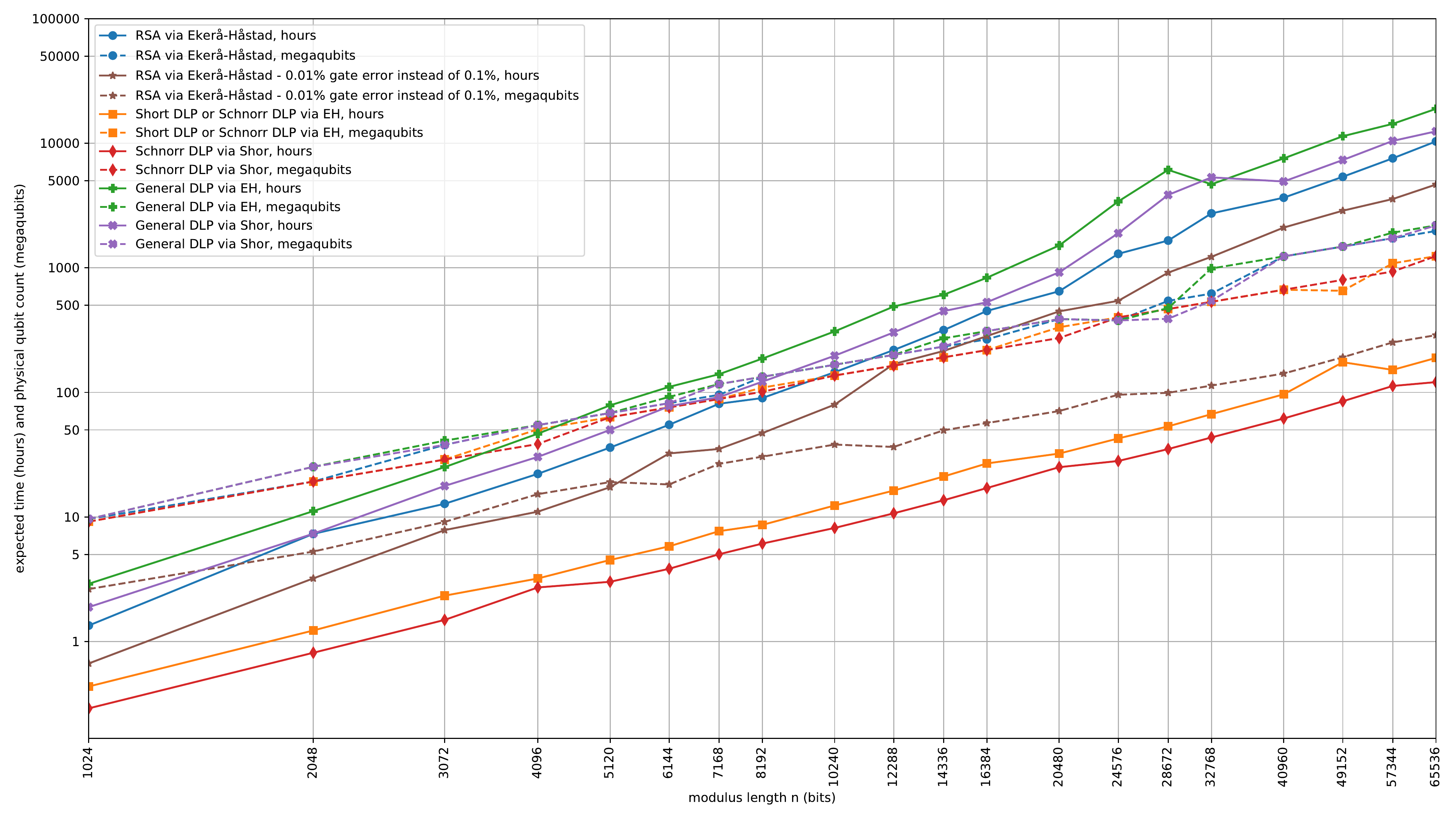}
    \end{center}
    \caption{
    Log-log plot of estimated space and expected-time costs, using our parallel construction, for various problems and problem sizes.
    See \sec{impact} for additional details.
    The jumps in space around $n=32786$ occur as the algorithm exceeds the error budget of the CCZ factory from \cite{gidney2018magic} and switches to the T factory from \cite{fowler2018}.
    Generated by ancillary file ``estimate\_costs.py".
    }
    \label{fig:plot-costs}
\end{figure}

\begin{table}[p]
  \resizebox{\textwidth}{!}{
  \begin{tabular}{r||c|c|c||c|c|c||c|c|c}
    &Abstract Qubits
        &Measurement Depth
            &Toffoli+T/2 Count
                &\multicolumn{3}{c||}{Toffoli+T/2 Count (billions)}
                    &\multicolumn{3}{c}{Min Volume (megaqubitdays)}
                        \\
\hline
Factoring RSA integers
    &\multicolumn{3}{c||}{Asymptotic}
        &$n=1024$ &$n=2048$ &$n=3072$
            &$n=1024$ &$n=2048$ &$n=3072$
                                \\
\hline
Vedral et al. 1996 \cite{vedral1996arithmetic}
    &$7n+1$
    &$80n^3 + O(n^2)$
    &$80n^3+O(n^2)$
        &86
        &690
        &2300
            &240
            &4100
            &23000
                \\
Zalka 1998 (basic) \cite{zalka1998fast}
    &$3n+O(1)$
    &$12n^3+O(n)$
    &$12n^3 + O(n^2)$
        &13
        &100
        &350
            &16
            &250
            &1400
                \\
Zalka 1998 (log add) \cite{zalka1998fast}
    &$5n+O(1)$
    &$600n^2+O(n)$
    &$52n^3 + O(n^2)$
        &56
        &450
        &1500
            &16
            &160
            &540
                \\
Zalka 1998 (fft mult) \cite{zalka1998fast}
    &$\approx 96n$
    &$\approx 2^{17} n^{1.2}$
    &$\approx 2^{17} n^2$
        &140
        &550
        &1200
            &62
            &260
            &710
                \\
Beauregard 2002 \cite{beauregard2002shor}
    &$2n+3$
    &$144 n^3 \lg n + O(n^2 \lg n)$
    &$576 n^3 \lg^2 n + O(n^3 \lg n)$
        &62000
        &600000
        &2200000
            &32000
            &380000
            &1700000
                \\
Fowler et al. 2012 \cite{fowler2012surfacecodereview}
    &$3n+O(1)$
    &$40n^3+O(n^2)$
    &$40n^3 + O(n^2)$
        &43
        &340
        &1200
            &53
            &850
            &4600
                \\
H\"{a}ner et al. 2016 \cite{haner2016factoring}
    &$2n+2$
    &$52n^3+O(n^2)$
    &$64n^3\lg n+O(n^3)$
        &580
        &5200
        &19000
            &230
            &2800
            &13000
                \\
\textbf{(ours) 2019}
    &{\boldmath $3n + 0.002 n \lg n$}
    &{\boldmath $500 n^2 + n^2 \lg n$}
    &{\boldmath $0.3 n^3 + 0.0005 n^3 \lg n$}
        &{\boldmath $0.4$}
        &{\boldmath $2.7$}
        &{\boldmath $9.9$}
            &{\boldmath $0.5$}
            &{\boldmath $5.9$}
            &{\boldmath $21$}
                \\
\hline
Solving elliptic curve DLPs
    &\multicolumn{3}{c||}{Asymptotic}
        &$n=160$ &$n=224$ &$n=256$
            &$n=160$ &$n=224$ &$n=256$
                \\
\hline
Roetteler et al. 2017 \cite{roetteler2017quantum}
    &$9n + O(\lg n)$
    &$448 n^3 \lg n + 4090 n^3$
    &$448 n^3 \lg n + 4090 n^3$
        &30
        &84
        &130
            &13
            &52
            &83
  \end{tabular}
  }
  \caption{
    Expected costs of factoring $n$ bit RSA integers using various constructions proposed in the literature.
    For comparison, we include a single construction for solving the DLP in $n$ bit prime order elliptic curve groups with comparable classical security levels.
    The estimated minimum spacetime volumes assume modern surface code constructions, even for older papers.
    See \app{table-details} for details on each entry in this table.
  }
  \label{tbl:comparison}
\end{table}

\begin{table}[p]
\resizebox{\linewidth}{!}{
  \begin{tabular}{r||c|c|c|c||c|c||c|c|c}
    &\multicolumn{4}{c||}{Physical assumptions}
    &\multicolumn{2}{c||}{Approach}
    &\multicolumn{3}{c}{Estimated costs}
                    \\
\hline
Historical cost
    &Physical gate
    &Cycle time
    &Reaction time
    &Physical
    &Distillation
    &Execution
    &Physical qubits
    &Expected runtime
    &Expected volume
        \\
estimate at $n=2048$
    &error rate
    &(microseconds)
    &(microseconds)
    &connectivity
    &strategy
    &strategy
    &(millions)
    &(days)
    &(megaqubitdays)
        \\
\hline
Van Meter et al. 2009 \cite{van2010distributed}
    &0.2\%
    &49
    &N/A
    &planar
    &1000+ T and S
    &distillation limited carry lookahead
    &6500
    &410
    &2600000
        \\
Jones et al. 2010 \cite{jones2012layered}
    &0.1\%
    &0.25
    &N/A
    &planar
    &7000 T
    &distillation limited carry lookahead
    &620
    &10
    &6200
        \\
Fowler et al. 2012 \cite{fowler2012surfacecodereview}
    &0.1\%
    &1
    &0.1
    &planar
    &1200 T
    &reaction limited ripple carry
    &1000
    &1.1
    &1100
        \\
O'Gorman et al. 2017 \cite{ogorman2017factories}
    &0.1\%
    &10
    &1
    &arbitrary
    &block CCZ
    &reaction limited ripple carry
    &230
    &3.7
    &850
        \\
Gheorghiu et al. 2019 \cite{gheorghiu2019cryptanalysis}
    &0.1\%
    &0.2
    &0.1
    &planar
    &1100 T
    &reaction limited ripple carry
    &170
    &1
    &170
        \\
(ours) 2019 (1 factory)
    &0.1\%
    &1
    &10
    &planar
    &1 CCZ
    &distillation limited ripple carry
    &16
    &6
    &90
        \\
(ours) 2019 (1 thread)
    &0.1\%
    &1
    &10
    &planar
    &14 CCZ
    &reaction limited ripple carry
    &19
    &0.36
    &6.6
        \\
\textbf{(ours) 2019 (parallel)}
    &\textbf{0.1\%}
    &\textbf{1}
    &\textbf{10}
    &\textbf{planar}
    &\textbf{28 CCZ}
    &\textbf{reaction limited carry runways}
    &\textbf{20}
    &\textbf{0.31}
    &\textbf{5.9}
        \\
  \end{tabular}
}
  \caption{
    Historical estimates of the expected costs of factoring $n=2048$ bit RSA integers, and the assumptions they used.
    We caution the reader to carefully consider the physical assumptions of each paper when comparing their expected volumes.
    For example, assuming a cycle time that is 5x more optimistic will reduce the expected volume by a corresponding factor of 5.
    See \app{historical-table-details} for details on each entry in this table.
  }
  \label{tbl:historical-comparison}
\end{table}

We use the standard square-and-multiply approach to reduce the exponentiations into a sequence of modular multiplications.
We apply optimizations to reduce the number of multiplications and the cost of each multiplication.

The number of multiplications is reduced by using windowed arithmetic \cite{van2005fastexponentiation,kutin2006shor,gidney2019windowedarithmetic}, which uses small table lookups to fuse several multiplications together.
It is also reduced by using Ekerå and Håstad's derivatives \cite{ekeraa2016modifying, ekeraa2017quantum, ekeraa2017pp, ekeraa2018general} of Shor's algorithms, that require fewer multiplications to be performed  compared to Shor's original algorithms.

The cost of each multiplication is reduced by combining several optimizations.
We use Zalka's coset representation of modular integers \cite{zalka2006pure}, which allows the use of cheaper non-modular arithmetic circuits to perform modular additions.
We use oblivious carry runways \cite{gidney2019approximatepermutation} to split registers into independent pieces that can be worked on in parallel when performing additions.
We bound the approximation error introduced by using oblivious carry runways and the coset representation of modular integers by analyzing them as approximate encoded permutations \cite{gidney2019approximatepermutation}.
We use windowed arithmetic (again) to fuse multiple additions into individual lookup additions \cite{gidney2019windowedarithmetic}.
We use a layout of the core lookup addition operation where carry propagation is limited by the reaction time of the classical control system, and where the lookups are nearly reaction limited \cite{gidney2019autoccz}.
Finally, we optimize the overall computation by trying many possible parameter sets (e.g. window sizes and code distances) for each problem size and selecting the best parameter set.

We estimate the approximate cost of our construction, both in the abstract circuit model, and in terms of its runtime and physical qubit usage in an error corrected implementation under plausible physical assumptions for large-scale superconducting qubit platforms with nearest neighbor connectivity (see \fig{plot-costs}).
Our physical estimates are based on the surface code \cite{fowler2012surfacecodereview}.
We assume distance-$d$ logical qubits are stored using square patches of $2(d+1)^2$ physical qubits (see \fig{lattice-surgery-qubit}) and are operated on via lattice surgery \cite{horsman2012latticesurgery,fowler2018}.
We provide concrete cost estimates for several cryptographically relevant problems, such as the RSA integer factoring problem, and various parameterizations of the DLP in finite fields.
These cost estimates may be used to inform decisions on when to mandate migration from currently deployed vulnerable cryptosystems to post-quantum secure systems or hybrid systems.

We caution the reader that, although we report two significant figures in our tables, there are large systemic uncertainties in our estimates.
For example, doubling (or halving) the physical gate error rate would increase (or decrease) the number of qubits required by more than $10\%$.
The estimates presented are intended to be ballpark figures that the cryptographic community can use to understand the potential impact of quantum computers; not exacting predictions of the future.

Compared to previous works, we reduce the Toffoli count when factoring RSA integers by over 10x (see \tbl{comparison}).
To only compare the Toffoli counts as in \tbl{comparison} may prove misleading, however, as it ignores the cost of routing, the benefits of parallelization, etc.
Ideally, we would like to compare our runtime and physical qubit usage to previous works in the literature.
However, this is only possible when such estimates are actually reported and use physical assumptions similar to our own.
The number of works for which this requirement is met is limited.

The works by Jones et al. \cite{jones2012layered}, Fowler et al. \cite{fowler2012surfacecodereview}, and Gheorgiu et al. \cite{gheorghiu2019cryptanalysis} stand out in that they use the same basic cost model as we use in this paper, enabling reasonable comparisons to be made.
We improve on their estimates by over 100x (see \tbl{historical-comparison}) when accounting for slight remaining differences in the cost model.
We also include other historical cost estimates \cite{van2010distributed,ogorman2017factories}, although in these cases the differences in physical assumptions are substantial.

As most previous works focus either on factoring, or on the DLP in elliptic curve groups, it is hard to find references against which to compare our cost estimates for solving the DLP in multiplicative groups of finite fields.
In general, the improvements we achieve for factoring RSA integers are comparable to the improvements we achieve for solving the general DLP in finite fields.
As may be seen in \fig{plot-costs}, the choice of parameterization has a significant impact on the costs.
For the short DLP, and the DLP in Schnorr groups, we achieve significant improvements.
These are primarily due to our use of derivatives of Shor's algorithm that are optimized for these parameterizations.

\subsection{Notation and conventions}

Throughout this paper, we refer to the modulus as $N$.
The modulus is the composite integer to be factored, or the prime characteristic of the finite field when computing discrete logarithms.
The number of bits in $N$ is $n = \lceil \lg N \rceil$ where $\lg(x) = \log_2(x)$.
The number of modular multiplication operations to perform (i.e. the combined exponent length in the modular exponentiations) is denoted $\lenexp$.
Our construction has a few adjustable parameters, which we refer to as $\gexp$ (the exponent window length), $\gmul$ (the multiplication window length), $\gsep$ (the runway separation), and $\gpad$ (the padding length used in approximate representations).

In the examples and figures, we consider moduli of length $n = 2048$ bits when $n$ needs to be explicitly specified.
This is because $n = 2048$ bits is the default modulus length in several widely used software programs \cite{ssh-keygen-man-page2018, gpg-faq-key-size2018, open-ssl-source2018}.
Our optimizations are not specific to this choice of $n$.
\sec{impact} provides cost estimates for a range of cryptographically relevant modulus lengths $n$.

We often quote costs as a function of both the number of exponent qubits $\lenexp$ and the problem size $n$.
We do this because the relationship between $\lenexp$ and $n$ changes from problem to problem, and optimizations that improve $\lenexp$ are orthogonal to optimizations that improve the cost of individual operations.

\subsection{On the structure of this paper}

The remainder of this paper is organized as follows:
In \sec{construction}, we describe our construction and the optimizations it uses, and show how to estimate its costs.
We then proceed in
\sec{impact} to describe how existing widely deployed cryptosystems are impacted by our work.
In \sec{future-work}, we present several ideas and possible optimizations that we believe are worth exploring in the future.
Finally, we summarize our contributions and their implications in \sec{conclusion}.

\section{Our construction}
\label{sec:construction}

\subsection{Quantum algorithms}

In Shor's original algorithm \cite{shor1994}, composite integers $N$ that are not pure powers are factored by computing the order~$r$ of a randomly selected element $\gen \in \mathbb Z_N^*$.
Specifically, period finding is performed with respect to the function $f(e) = \gen^e$.
This involves computing a modular exponentiation with $\lenexp=2n$ qubits in the exponent $e$.

If $r$ is even and $\gen^{r/2} \not\equiv -1$, which holds with probability at least $1/2$ by~\cite{shor1994}, this yields non-trivial factors of $N$.
To see why, lift $\gen$ to $\mathbb Z$.
As $\gen^r - 1 = (\gen^{r/2} - 1)(\gen^{r/2} + 1)
\equiv 0 \pmod{N}$ it suffices to compute $\gcd((\gen^{r/2} \pm 1) \text{ mod } N, N)$ to factor $N$.

In \cite{ekeraa2016modifying,ekeraa2017quantum,ekeraa2017pp}, Ekerå and Håstad explain how to factor RSA integers $N = pq$ in a different way; namely by computing a short discrete logarithm.
This algorithm proceeds as follows:
First $y = \gen^{N+1}$ is computed classically, where as before~$\gen$ is randomly selected from $\mathbb Z_N^*$ and of unknown order~$r$.
Then the discrete logarithm $d = \log_{\gen} y \equiv p + q \pmod{r}$ is computed on the quantum computer.
To see why $d \equiv p + q \pmod{r}$, note that $\mathbb Z_N^*$ has order $\phi(N) = (p-1)(q-1)$ by Euler's totient theorem, so $d \equiv pq + 1 \equiv p + q \pmod{r}$ as $r$ divides $\phi(N)$ and $pq + 1 \equiv p + q \pmod{\phi(N)}$, with equality if $r > p + q$.
For large random RSA integers, the order $r > p + q$ with overwhelming probability.
Hence, we may assume $d = p + q$.
By using that $N = pq$ and $d=p+q$, where~$N$ and~$d$ are both known, it is trivial to deterministically recover the factors $p$ and $q$ as the roots of the quadratic equation $p^2 - dp + N = 0$.

The quantum part of Ekerå and Håstad's algorithm is similar to the quantum part of Shor's algorithm, except for the following important differences:
there are two exponents $e_1, e_2$, of lengths $2m$ and $m$ qubits respectively.
The value $m$ is a positive integer that must satisfy $p + q < 2^m$.
We set $m = 0.5n + O(1)$.
Period finding is performed against the function $f(e_1, e_2) = \gen^{e_1} y^{-e_2}$ rather than the function $f(e)=\gen^e$.
The total exponent length is hence $n_e = 3m = 1.5n + O(1)$ qubits, compared to $2n$ qubits in Shor's algorithm.
It is this reduction in the exponent length that translates into a reduction in the overall number of multiplications that are performed quantumly.

The two exponent registers are initialized to uniform superpositions of all $2^{2m}$ and $2^{m}$ values, respectively, and two quantum Fourier transforms are applied independently to the two registers.
This implies that standard optimization techniques, such as the square-and-multiply technique, the semi-classical Fourier transform of Griffiths and Niu \cite{griffiths1996semiclassical}, recycling of qubits in the exponent registers \cite{mosca1999recycle, parker2000recycle}, and so forth, are directly applicable.

The classical post-processing algorithm used to extract the logarithm from the observed frequencies is by necessity different from Shor's.
It uses lattice-based techniques, and critically does not require $r$ to be known.
Ekerå shows in~\cite{ekeraa2017pp} that the post-processing algorithm has probability above $99\%$ of recovering $d$.
As the factors $p$ and $q$ are recovered deterministically from $d$ and $N$, this implies that it in general suffices to run the quantum algorithm once.

In summary, Ekerå and Håstad's algorithm is similar to Shor's algorithm from an implementation perspective:
a sequence of multiplications is computed, where one operand is in a quantum register and one operand is a classical constant.
The multiplications are interleaved with the Fourier transform.
It is of no significance to our construction what sequence of classical constants are used, or if one or more independent Fourier transforms need to be applied.
This fact, coupled with the fact that the algorithms of Ekerå and Håstad may be used to solve other relevant problems where $n_e$ is a different function of $n$ (see \sec{impact}), lead us to describe our construction in terms of $n$ and $n_e$.

Note that the above description of the algorithm has been somewhat simplified compared to the algorithm described in \cite{ekeraa2017quantum, ekeraa2017pp} in the interest of improved legibility.
Furthermore, there are technical conditions that need to be respected for the analysis in \cite{ekeraa2017pp} to be applicable.
For the full details, see Appendix~A.2.1 of \cite{ekeraa2017pp}.

\subsection{Reference implementation}

To avoid overloading the reader, we will describe our factoring construction by starting from a simple reference implementation of the quantum part of Shor's original algorithm and then apply optimizations one by one.

The reference implementation works the way most implementations of Shor's algorithm do, by decomposing exponentiation into iterative controlled modular multiplication \cite{vedral1996arithmetic, zalka1998fast, zalka2006pure, beauregard2002shor, haner2016factoring, gidney2017factoring}.
A register $\productreg$ is initialized to the $|1\rangle$ state, then a controlled modular multiplication of the classical constant $\gen^{2^j} \pmod{N}$ into $\productreg$ is performed, controlled by the qubit $e_j$ from the exponent $e$, for each integer $j$ from $\lenexp-1$ down to $0$.
After the multiplications are done, $\productreg$ is storing $\gen^e \pmod{N}$ and measuring $\productreg$ completes the hard part of Shor's algorithm.

Controlled modular multiplication is still not a primitive operation, so it must also be decomposed.
It can be performed by introducing a work register $\workreg$ initialized to $|0\rangle$ and then performing the following two controlled scaled additions: $\workreg \mathrel{+}= \productreg \cdot k \pmod{N}$ then $\productreg \mathrel{+}= \workreg \cdot (-k^{-1}) \pmod{N}$.
After these two operations, $\workreg$ is storing the result and $\productreg$ has been cleared to the $|0\rangle$ state.
Swapping the two registers, so that the result is in $\productreg$, completes the multiplication.

Performing a controlled scaled addition with classical scale factor $k$ can be done with a series of~$n$ controlled modular additions.
For each qubit $q_j$ in the input register, you add $k \cdot 2^j \pmod{N}$ into the target, controlled by $q_j$.
Controlled modular addition in turn is performed via a series of non-modular additions and comparisons.
For example, \cite{vedral1996arithmetic} uses five additions for this purpose.
Finally, using the Cuccaro adder \cite{cuccaro2004adder}, uncontrolled non-modular additions can be performed with~$O(1)$ additional workspace using $2n$ Toffolis.

Combining the numbers in the preceding two paragraphs implies a Toffoli count of $\lenexp \cdot 2n \cdot 5 \cdot 2n = 20 \lenexp n^2$ for the reference implementation.
It constitutes the baseline for the optimizations that we apply in the following sections.

\subsection{The quantum Fourier transform}
Before proceeding, note that we focus exclusively on costing the controlled modular multiplications, both in the reference implementation and throughout the remainder of this paper.
This is because they constitute the hard part of Shor's algorithm from an implementation perspective.
The other part of the algorithm, the quantum Fourier transform (QFT), can be implemented semi-classically~\cite{griffiths1996semiclassical} and interleaved with the controlled modular multiplications.

When the QFT is implemented in this manner (see~\fig{exponent-grouping}) it suffices to have control qubits for the window of controlled modular multiplications currently being processed, as these control qubits may be recycled \cite{mosca1999recycle, parker2000recycle}. In the QFT, a sequence of classically controlled gates that shift the phase by $2\pi/2^k$ for various integers $k$ up to the QFT dimension are applied to the control qubit for each modular multiplication.
This sequence can be combined into a single shift, and implemented using e.g. the technique in \cite{bocharov2015rus}, to a level of precision far below other sources of error in the algorithm without incurring a noticeable cost. This implies that the cost of implementing the QFT is negligible.

Note furthermore that when we analyze the success probabilities of Shor's algorithms, and the various derivatives, we assume the use of an ideal QFT even though the implemented QFT is technically an approximation.

\subsection{The coset representation of modular integers}

Following Zalka \cite{zalka2006pure}, the first improvement we make over the reference implementation is to switch from the usual representation of integers to the coset representation of modular integers.
The usual representation of an integer $k$ in a quantum computer is the computational basis state $|k\rangle$.

The coset representation is different: It uses a periodic superposition with period~$N$ and offset~$k$.
In ideal conditions, the integer $k \pmod{N}$ is represented by the state $\sqrt{2^{-\gpad}} \sum_{j=0}^{2^\gpad-1} |jN + k\rangle$ where $\gpad$ is the number of additional padding qubits placed at the end of the register after the high order bits.
The key idea is that the periodicity of the superposition causes a non-modular adder to perform approximate modular addition on this representation, and the error of the approximation can be exponentially suppressed by adding more padding to the register.

As we will discuss later, the amount of padding required is logarithmic in the number of operations.
This small cost enables the large benefit of using non-modular adders instead of modular adders.
It is possible to perform a controlled non-modular addition in $4n$ Toffolis \cite{cuccaro2004adder, gidney2018addition}, significantly cheaper than the $10n$ we assumed for a controlled modular adder.
Therefore switching to the coset representation reduces the leading asymptotic term of the Toffoli count of the reference implementation from $20 \lenexp n^2$ to $8 \lenexp n^2$.

\subsection{Windowed arithmetic}

The next optimization we use is windowed arithmetic \cite{van2005fastexponentiation,kutin2006shor,gidney2019windowedarithmetic}.
Specifically, we follow the modular exponentiation construction from \cite{gidney2019windowedarithmetic} and use windowing at two levels.

First, at the level of a multiplication, we window over the controlled additions.
We fuse groups of controlled additions into single lookup additions.
A lookup addition is an addition where the value to add into a register is the result of a table lookup.
Small windows of the qubits that would have been used as controls for the additions are instead treated as addresses into classically precomputed tables of values to unconditionally add into the target.

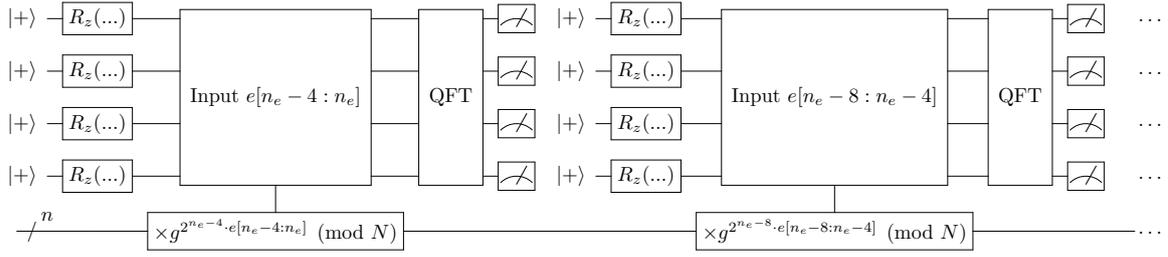
\begin{figure}
\resizebox{\textwidth}{!}{
\Qcircuit @R=1em @C=0.75em {
\\
&&\lstick{|+\rangle} &\gate{R_z(...)} &       \multigate{3}{\text{Input }e[\lenexp-4:\lenexp]}&       \multigate{3}{\text{QFT}}&\meter&&&&\lstick{|+\rangle} &\gate{R_z(...)} &       \multigate{3}{\text{Input }e[\lenexp-8:\lenexp-4]}&       \multigate{3}{\text{QFT}}&\meter&&&\dots\\
&&\lstick{|+\rangle} &\gate{R_z(...)} &              \ghost{\text{Input }e[\lenexp-4:\lenexp]}&              \ghost{\text{QFT}}&\meter&&&&\lstick{|+\rangle} &\gate{R_z(...)} &              \ghost{\text{Input }e[\lenexp-8:\lenexp-4]}&              \ghost{\text{QFT}}&\meter&&&\dots\\
&&\lstick{|+\rangle} &\gate{R_z(...)} &              \ghost{\text{Input }e[\lenexp-4:\lenexp]}&              \ghost{\text{QFT}}&\meter&&&&\lstick{|+\rangle} &\gate{R_z(...)} &              \ghost{\text{Input }e[\lenexp-8:\lenexp-4]}&              \ghost{\text{QFT}}&\meter&&&\dots\\
&&\lstick{|+\rangle} &\gate{R_z(...)} &              \ghost{\text{Input }e[\lenexp-4:\lenexp]}&              \ghost{\text{QFT}}&\meter&&&&\lstick{|+\rangle} &\gate{R_z(...)} &              \ghost{\text{Input }e[\lenexp-8:\lenexp-4]}&              \ghost{\text{QFT}}&\meter&&&\dots\\
&{/}\qw&\ustick{n}\qw&\qw             &\gate{\times \gen^{2^{n_e-4} \cdot e[\lenexp-4:\lenexp]} \pmod{N}}\qwx&                      \qw&   \qw&\qw&\qw&\qw&\qw       &\qw             &\gate{\times \gen^{2^{n_e-8} \cdot e[\lenexp-8:\lenexp-4]} \pmod{N}}\qwx&                      \qw&\qw&\qw&\qw&\dots\\
\\
}
}
    \caption{
    Working through the qubits representing an exponent in Shor's algorithm with a window size $\gexp$ of four, while using a semi-classical Fourier transform \cite{griffiths1996semiclassical} with recycling \cite{mosca1999recycle, parker2000recycle}.
    The notation $e[a:b] = \lfloor e / 2^a \rfloor \bmod 2^{b-a}$ refers to a slice of (qu)bits in $e$ from (qu)bit index $a$ (inclusive) to (qu)bit index $b$ (exclusive).
    Each $R_z(\dots)$ gate rotates by an amount determined by previous measurements, e.g. using the technique in \cite{bocharov2015rus}.
    All 16 possible values of the expression $\gen^{e[4:8] 2^4} \pmod{N}$ (and similar expressions) can be precomputed classically, and looked up on demand within the multiplication circuit.
    This reduces the number of multiplications by a factor of the window size, at the cost of some additional lookup work within the multiplication.
    }
    \label{fig:exponent-grouping}
\end{figure}

Second, at the level of the exponentiation, we window over the controlled multiplications.
This is done by including exponent qubits in the addresses of all lookups being performed within the multiplications.
We refer the reader to~\cite{gidney2019windowedarithmetic} for the exact details of this nested windowed arithmetic construction.

The cost of windowed arithmetic depends on the size of the windows.
Let $\gexp$ be the size of the window over exponent qubits that are being used to control multiplications.
Increasing $\gexp$ proportionally decreases the number of multiplications needed for the modular exponentiation, since more exponent qubits are handled by each multiplication.
Let $\gmul$ be the size of the window over factor qubits being used to control additions.
Increasing $\gmul$ proportionally decreases the number of additions needed within each multiplication.
By using windowed arithmetic, the $\lenexp$ controlled multiplications we needed to perform become $\lenexp / \gexp$ uncontrolled multiplications while the $2 n$ controlled additions we needed to perform within each multiplication become $2 n / \gmul$ uncontrolled additions.
The cost of increasing $\gexp$ and $\gmul$ is that each addition is now accompanied by a lookup, and the Toffoli count of the lookup operation is $2^{\gexp + \gmul}$.

Using Cuccaro et al.'s adder \cite{cuccaro2004adder}, each $n$-bit addition has a Toffoli count and measurement depth of $2n$.
Using Babbush et al.'s QROM read (section 3A of \cite{babbush2018}), each table lookup has a Toffoli count and measurement depth of $2^{\gmul+\gexp}$ and uses a negligible $O(\gmul + \gexp)$ ancillae.
Using measurement-based uncomputation (appendix~C of \cite{berry2019qubitization}), uncomputing each table lookup has a Toffoli count and measurement depth of $2 \sqrt{2^{\gmul + \gexp}}$ which is negligible compared to the lookup for the parameters we will be using.

The measurement-based uncomputation uses $\text{max}(0, \sqrt{2^{\gmul + \gexp}} - n)$ ancillae (it starts by measuring away $n$ logical qubits of the lookup output register).
This ancillae count is always zero for the reasonable values of $\gmul, \gexp$, and classically intractable values of $n$, that we consider in this paper.
The overhead due to the logarithmic padding introduced by the coset representation of modular integers is logarithmic in $n$.

Thus, by using windowed arithmetic, the leading term of the Toffoli count of the reference implementation has been reduced to $\frac{2 \lenexp n}{\gmul \gexp} (2n + 2^{\gmul + \gexp})$.
If we set $\gmul = \gexp = \frac{1}{2} \lg n$ then we achieve a Toffoli count with leading term $\frac{24 \lenexp n^2}{\lg^2 n}$.
In terms of space, we are using $3n + O(\lg n)$ logical qubits.
The qubits are distributed as follows:
  $n+O(\lg n)$ for the accumulation register storing the product,
  $n+O(\lg n)$ for a workspace register during the multiplication,
  $n + O(\lg n)$ for the lookup output register,
  and $O(\lg n)$ qubits to hold the part of the exponent needed for the current stage of the semi-classical Fourier transform \cite{griffiths1996semiclassical}.

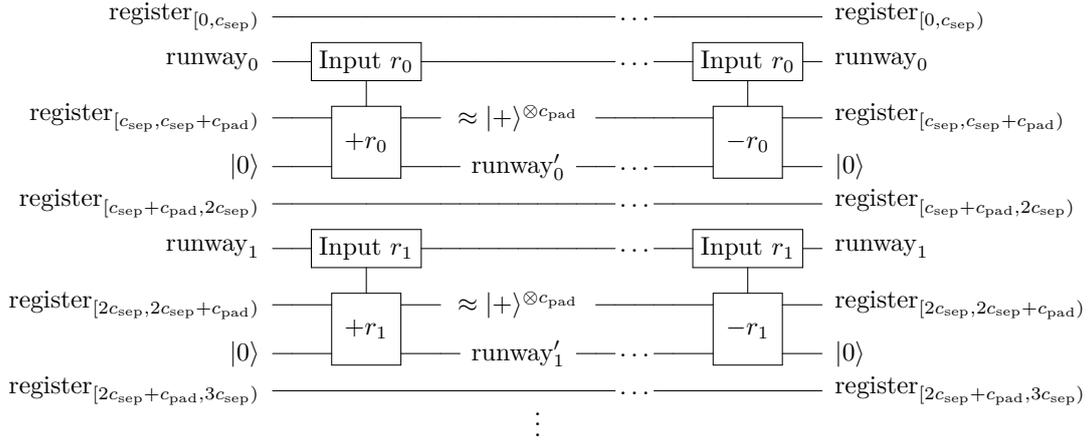
\begin{figure}
  \resizebox{\textwidth}{!}{
  \Qcircuit @R=1em @C=0.75em {
  \\
  &&&&&&&&&&&&&&&\lstick{\text{register}_{[0, \gsep)}}            &\qw&\qw                     &\qw&\qw&\qw&\qw&\qw                            &\qw&\qw&\qw&\qw&\qw&\dots&   &\qw &\qw                     &\rstick{\text{register}_{[0, \gsep)}}           \qw&&&&&&&&&&&&&&&\\
  &&&&&&&&&&&&&&&\lstick{\text{runway}_0}                         &\qw&\gate{\text{Input }r_0} &\qw&\qw&\qw&\qw&\qw                            &\qw&\qw&\qw&\qw&\qw&\dots&   &\qw &\gate{\text{Input }r_0} &\rstick{\text{runway}_0}                        \qw&&&&&&&&&&&&&&&\\
  &&&&&&&&&&&&&&&\lstick{\text{register}_{[\gsep,\gsep+\gpad)}}   &\qw&\multigate{1}{+r_0} \qwx&\qw&   &   &   &\approx|+\rangle^{\otimes\gpad}&   &   &   &   &\qw&\qw  &\qw&\qw &\multigate{1}{-r_0} \qwx&\rstick{\text{register}_{[\gsep,\gsep+\gpad)}}  \qw&&&&&&&&&&&&&&&\\
  &&&&&&&&&&&&&&&\lstick{|0\rangle}                               &\qw&\ghost{+r_0}            &\qw&\qw&   &   &\text{runway}_0^\prime         &   &   &   &\qw&\qw&\dots&   &\qw &\ghost{-r_0}            &\rstick{|0\rangle}                              \qw&&&&&&&&&&&&&&&\\
  &&&&&&&&&&&&&&&\lstick{\text{register}_{[\gsep+\gpad,2\gsep)}}  &\qw&\qw                     &\qw&\qw&\qw&\qw&\qw                            &\qw&\qw&\qw&\qw&\qw&\dots&   &\qw &\qw                     &\rstick{\text{register}_{[\gsep+\gpad,2\gsep)}} \qw&&&&&&&&&&&&&&&\\
  &&&&&&&&&&&&&&&\lstick{\text{runway}_1}                         &\qw&\gate{\text{Input }r_1} &\qw&\qw&\qw&\qw&\qw                            &\qw&\qw&\qw&\qw&\qw&\dots&   &\qw &\gate{\text{Input }r_1} &\rstick{\text{runway}_1}                        \qw&&&&&&&&&&&&&&&\\
  &&&&&&&&&&&&&&&\lstick{\text{register}_{[2\gsep,2\gsep+\gpad)}} &\qw&\multigate{1}{+r_1} \qwx&\qw&   &   &   &\approx|+\rangle^{\otimes\gpad}&   &   &   &   &\qw&\qw  &\qw&\qw &\multigate{1}{-r_1} \qwx&\rstick{\text{register}_{[2\gsep,2\gsep+\gpad)}}\qw&&&&&&&&&&&&&&&\\
  &&&&&&&&&&&&&&&\lstick{|0\rangle}                               &\qw&\ghost{+r_1}            &\qw&\qw&   &   &\text{runway}_1^\prime         &   &   &   &\qw&\qw&\dots&   &\qw &\ghost{-r_1}            &\rstick{|0\rangle}                              \qw&&&&&&&&&&&&&&&\\
  &&&&&&&&&&&&&&&\lstick{\text{register}_{[2\gsep+\gpad,3\gsep)}} &\qw&\qw                     &\qw&\qw&\qw&\qw&\qw                            &\qw&\qw&\qw&\qw&\qw&\dots&   &\qw &\qw                     &\rstick{\text{register}_{[2\gsep+\gpad,3\gsep)}}\qw&&&&&&&&&&&&&&&\\
  &&&&&&&&&&&&&&&&&&&&&&&\vdots\\
  \\
  }
  }
  \caption{
    How to temporarily reduce oblivious carry runways to a single qubit, in preparation for being used as the input factor in a multiply-add operation that must iterate over all qubits in the register.
    The multiply-add should occur during the ``$\dots$" section in the middle.
  }
  \label{fig:temp-register-fold}
\end{figure}

\subsection{Oblivious carry runways}

The last major algorithmic optimization we apply is the use of oblivious carry runways \cite{gidney2019approximatepermutation}.
The basic problem addressed by runways is that, normally, a carry signal generated at the bottom of a register must propagate all the way to the top of the register.
This process can be short-circuited by instead terminating the carry propagation into an appropriately constructed runway.
Runways allow large additions to be performed piecewise, with each piece being worked on in parallel, by terminating carries into appropriately placed runways at the end of each piece.

As with the coset representation of modular integers, circuits using oblivious carry runways approximate the original circuit instead of perfectly reproducing it.
But, as we will discuss later, increasing the runway length $\gpad$ exponentially suppresses the approximation error.

For the full details of how to add, maintain, and remove oblivious runways we refer the reader to \cite{gidney2019approximatepermutation}.
That being said, we will point out one optimization not mentioned in that paper which applies here.
Note that, ultimately, each register we add carry runways to is going to either be measured or discarded.
The last thing that would happen to these registers, before they were measured or discarded, is carry runway removal.
However, the carry runway removal process is classical; it can be constructed out of Toffoli gates.
As a result, the carry runway removal process does not need to be performed under superposition.
The register qubits and runway qubits can simply be measured just before the runways would have been removed, and the runway removal process can be performed by classical post-processing of the measurements.

Note that we use $\gpad$ for both the runway length of oblivious carry runways and the padding length of the coset representation.
We do this because they have identical error mechanisms, namely unwanted carries into the extra qubits.
There is negligible benefit in suppressing one more than the other.

The major benefit of oblivious carry runways, compared to previous techniques for reducing the depth of addition such as Draper et al.'s logarithmic depth carry-lookahead adder \cite{draper2004logarithmic}, is that oblivious carry runways can be introduced gradually without incurring large overheads.
The Toffoli count and workspace overheads are linear in the number of pieces $\lceil n/\gsep\rceil$ (where $\gsep$ is the runway spacing) but only logarithmic in $n$.

For example, if you place a single runway at the midpoint of a 2048-qubit register, then the number of qubits and the number of Toffolis needed to perform an addition will increase by a couple percent but the depth of the additions is nearly halved.

\subsection{Interactions between optimizations}

A major benefit of the set of optimizations we have chosen to use in this paper is that they complement each other.
They make orthogonal improvements that compound or reinforce, instead of conflicting.

For example, when using the coset representation of modular integers, it is important that additions only use offsets less than $N$.
Larger offsets cause larger approximation error.
However, because we are using windowed arithmetic, every addition we perform has an offset that is being returned by a table lookup.
Since the entries in the tables are classically precomputed, we can classically ensure all offsets are canonicalized into the $[0, N)$ range.

There are two cases where the optimizations do not mesh perfectly.

First, using oblivious runways reduces the depth of addition but not the depth of lookups.
This changes their relative costs, which affects the optimal window sizes to use in windowed arithmetic.
When addition is suddenly four times faster than it used to be, the optimal window sizes~$\gexp$ and~$\gmul$ decrease by~$1$ (approximately).

Second, when iterating over input qubits during a multiply-add, it is necessary to also iterate over runway qubits and padding qubits.
This increases the number of lookup additions that need to be performed in order to complete a windowed multiplication.
This issue is partially solved by temporarily adding the runways back into the main register before performing a multiply-add where that register is used as one of the factors (instead of the target).
This temporarily reduces each runway to a single carry qubit (see \fig{temp-register-fold}).
We could reduce the runways to zero qubits, but this would require propagating the carry qubits all the way across the register, so we do not.

\subsection{Abstract circuit model cost estimate}

We have now described all the details necessary to estimate the cost of our implementation in the abstract circuit model.
The cost depends on the two parameters specified by the problem (the input size $n$ and the number of exponent qubits $\lenexp$) and also the four controllable parameters we have discussed (the exponentiation window size~$\gexp$, the multiplication window size $\gmul$, the runway separation $\gsep$, and the padding/runway length $\gpad$).

Although we do consider alternate values when producing tables and figures, in general we have found that the settings $\gexp=\gmul=5, \gsep=1024, \gpad=2 \lg n + \lg \lenexp + 10 \approx 3 \lg n + 10$ work well.
In this overview we will focus on these simple, though slightly suboptimal, values.

Recall that an exponentiation may be reduced to a sequence of $\lenexp$ multiplications, which we process in groups of size $\gexp$.
For each multiplication, we do two multiply-adds.
Each multiply-add will use several small additions to temporarily reduce the runway registers to single qubits.
The multiply-add then needs to perform a sequence of additions controlled by the $n$ main register qubits, the $O(\lg n)$ coset padding qubits, and the $n/\gsep$ reduced runway qubits.
Using windowed arithmetic, these additions are done in groups of size $\gmul$ with each group handled by one lookup addition.
Therefore the total number of lookup additions we need to perform is

\begin{equation}
\begin{aligned}
    \text{LookupAdditionCount}(n, \lenexp)
    &= \frac{2 n \lenexp}{\gexp \gmul} \cdot \frac{\gsep + 1}{\gsep} + O\left( \frac{\lenexp \lg n}{\gexp \gmul} \right)
    \\&\rightarrow \frac{2 n \lenexp}{25} \cdot \frac{1025}{1024} + O\left(\lenexp \lg n\right)
    \\&\approx 0.1 \lenexp n
\end{aligned}
\end{equation}

The lookup additions make up essentially the entire cost of the algorithm, i.e.
the total Toffoli count is approximately equal to the Toffoli count of a lookup addition times the lookup addition count.
This works similarly for the measurement depth.
Ignoring the negligible cost of uncomputing the lookup, the Toffoli count of a lookup addition is $2n + n \gpad / \gsep + 2^{\gexp + \gpad}$ and the measurement depth is $2 \gsep + 2 \gpad + 2^{\gexp + \gmul}$.
Therefore

\begin{equation}
\begin{aligned}
    \text{ToffoliCount}(n, \lenexp)
    &\approx \text{LookupAdditionCount}(n, \lenexp) \cdot \left(2n + \gpad \frac{n}{\gsep} + 2^{\gexp + \gpad} \right)
    \\&\approx 0.2 \lenexp n^2 + 0.0003 \lenexp n^2 \lg n
\end{aligned}
\end{equation}

\begin{equation}
\begin{aligned}
    \text{MeasurementDepth}(n, \lenexp)
    &\approx \text{LookupAdditionCount}(n, \lenexp) \cdot \left(2 \gsep + 2\gpad + 2^{\gexp + \gmul} \right)
    \\&\approx 300 \lenexp n + 0.5 \lenexp n \lg n
\end{aligned}
\end{equation}

These approximate upper bounds, with $\lenexp$ set to $1.5n$, are the formulae we report in the abstract and in \tbl{comparison} for the cost of factoring RSA integers.

\subsection{Approximation error}

Because we are using oblivious carry runways and the coset representation of modular integers, the computations we perform are not exact.
Rather, they are approximations.
This is not a problem in practice, as we may bound the approximation error using concepts and results from \cite{gidney2019approximatepermutation}.

The oblivious runways and the coset representation of modular integers are both examples of ``approximate encoded permutations" which have a ``deviation".
When using a padding/runway length of $\gpad$, and a runway separation of~$\gsep$, the deviation per addition is at most $n / (\gsep 2^{\gpad})$.
Deviation is subadditive under composition, so the deviation of the entire modular exponentiation process is at most the number of lookup additions times the deviation per addition.

We can use this to check how much deviation results from using $\gpad = 2 \lg n + \lg \lenexp + 10$, which we so far have merely claimed would be sufficient:

\begin{equation}
\begin{aligned}
    \text{TotalDeviation}(n, \lenexp)
    &\leq \text{LookupAdditionCount}(n, \lenexp) \cdot \frac{n}{\gsep 2^{\gpad}}
    \\&\approx 0.1 \lenexp n \frac{n}{1024 \cdot 2^{2 \lg n + \lg \lenexp + 10}}
    \\&\approx 0.1 \frac{n^2 \lenexp}{1024 \cdot n^2 \lenexp \cdot 2^{10}}
    \\&= 0.1 \frac{1}{1024 \cdot 2^{10}}
    \\&\approx 10^{-7}
\end{aligned}
\end{equation}

When an approximate encoded permutation has a deviation of $\epsilon$, the trace distance between its output and the ideal output is at most $2 \sqrt{\epsilon}$.
Therefore the approximation error (i.e. the trace distance due to approximations), using the parameter assignments we have described, is roughly~$0.1\%$.
This is significantly lower than other error rates we will calculate.

\begin{figure}
    \begin{center}
    \resizebox{0.9 \linewidth}{!}{
        \includegraphics{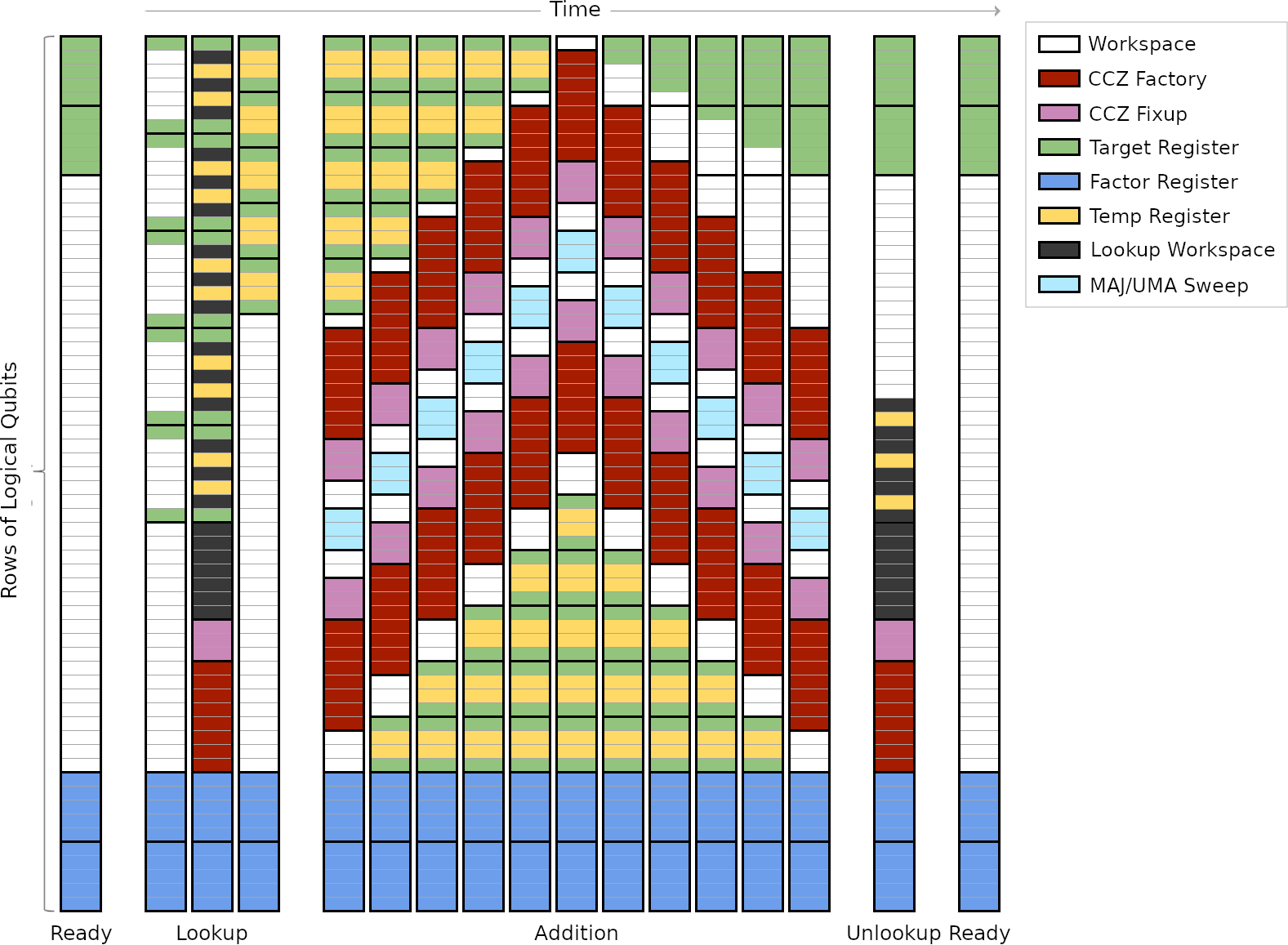}
    }
    \end{center}
    \caption{
        Space usage during the key lookup-add-unlookup inner loop of our construction.
        During lookup, the target register is spread out to make room for the temporary register that will hold the lookup's output.
        During addition the target register and lookup output register are squeezed through a moving operating area that sweeps up then down, applying the MAJ and UMA sweeps of Cuccaro's adder.
        Uncomputing the lookup is done with measurement-based uncomputation \cite{berry2019qubitization}, which is overlapped slightly with the UMA sweep (this is why the yellow rows disappear early).
        ``CCZ factory" space is being used to produce CCZ states.
        ``CCZ fixup" space is being used to convert the CCZ states into AutoCCZ states \cite{gidney2019autoccz}.
        ``MAJ/UMA" space is being used to perform a ripple carry addition, fueled by AutoCCZ states, of the temporary register (the result of the lookup) into the target register.
    }
    \label{fig:time-bars}
\end{figure}

\subsection{Spacetime layout}

Our implementation of Shor's algorithm uses a layout that is derived from the (nearly) reaction limited layouts presented in \cite{gidney2019autoccz}.
In these layouts, a lookup addition is performed as follows (see \fig{time-bars}).
All data qubits are arranged into rows.
The rows of qubits making up the target register of the addition are spread out into row pairs with gaps five rows wide in between.
In these gaps there will be two rows for the lookup output register, and three rows for access hallways.
This arrangement allows the lookup computation two ways to access each output row, doubling the speed at which it can run, while simultaneously ensuring the target register and the lookup output register are interleaved in a convenient fashion.

After the lookup is completed, the rows of the lookup output register and the addition target register are packed tightly against the top (or equivalently bottom) of the available area.
An operating area is prepared below them, and the data is gradually streamed through the operating area while the operating area gradually shifts upward.
This performs the MAJ sweep of Cuccaro's ripple carry adder \cite{cuccaro2004adder}.
Everything then begins moving in the opposite direction in order to perform the UMA sweep on the return stroke.

The lookup register is quickly uncomputed using measurement-based uncomputation \cite{berry2019qubitization}, and the system is returned to a state where another lookup addition can be performed.

\begin{figure}[p]
    \begin{center}
    \resizebox{0.94\linewidth}{!}{
        \includegraphics{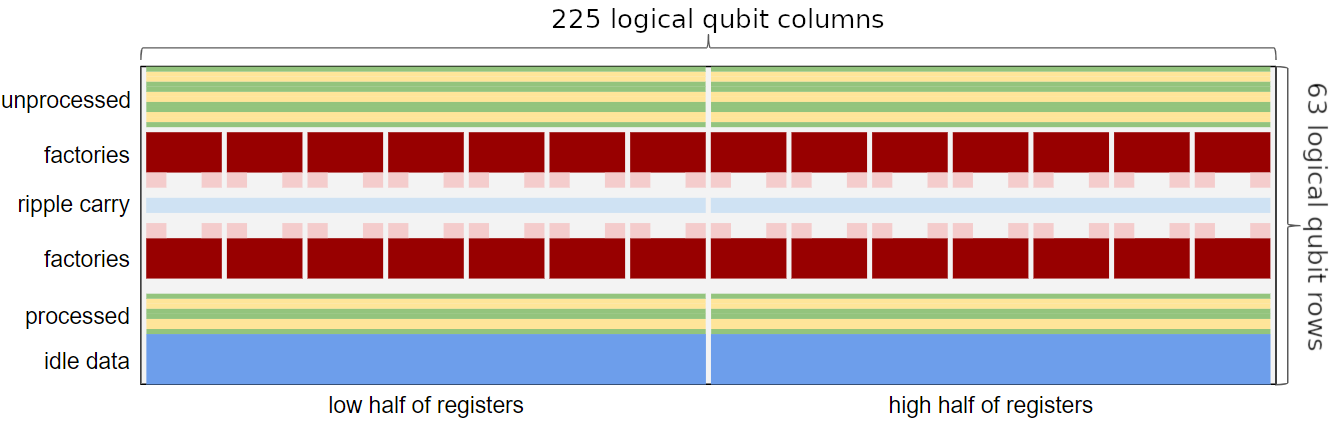}
    }
    \end{center}
    \caption{
        Data layout during a parallel addition.
        Corresponds to one of the ``addition" columns from \fig{time-bars}.
        To scale, assuming a 2048 bit number is being factored.
        The left and right halves of the computer run completely independently and in parallel.
        The factories (red boxes and pink boxes) are feeding AutoCCZ states into the blue area, which is rippling the carry bit back and forth as the offset and target register data (green and yellow rows) is routed through gaps between the factories to the other side.
    }
    \label{fig:addition-layout-2d}
\end{figure}

\begin{figure}[p]
    \begin{center}
    \resizebox{0.94\linewidth}{!}{
        \includegraphics{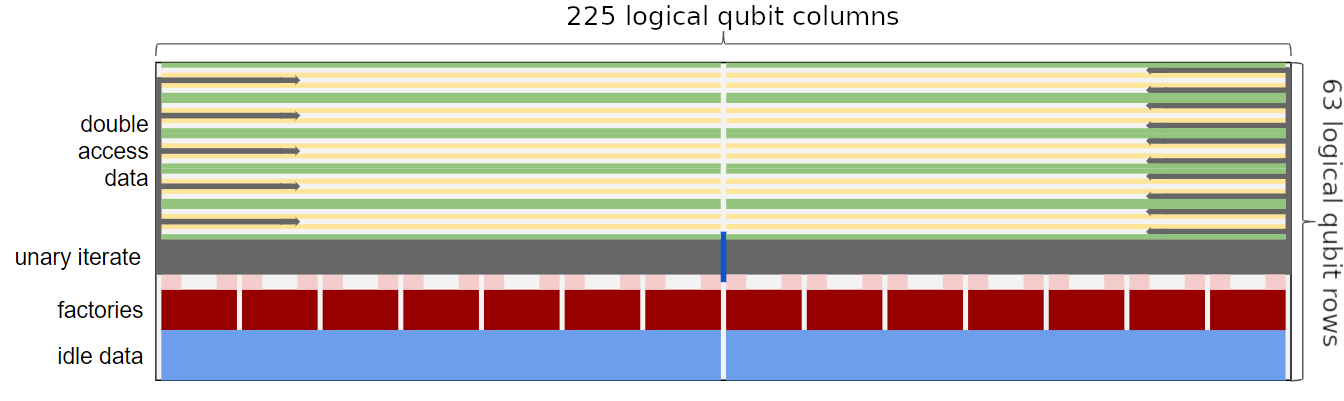}
    }
    \end{center}
    \caption{
        Data layout during a table lookup, as generalized from \cite{gidney2019autoccz}.
        To scale, assuming a 2048 bit number is being factored.
        Corresponds to the center ``lookup" column from \fig{time-bars}.
        The factories (red boxes and pink boxes) are feeding AutoCCZ states into the dark gray region, which is performing the unary iteration part of a table lookup computation.
        There are enough factories, and enough work space, to run the lookup at double speed, made possible by the fact that every qubit in the lookup output register (yellow) is adjacent to two access hallways.
        The target register (green rows) and factor register (blue rows) are idle, except that a few qubits from the factor register are being used as address bits in the table lookup.
    }
    \label{fig:lookup-layout}
\end{figure}

\begin{figure}[p]
    \begin{center}
    \resizebox{\linewidth}{!}{
        \includegraphics{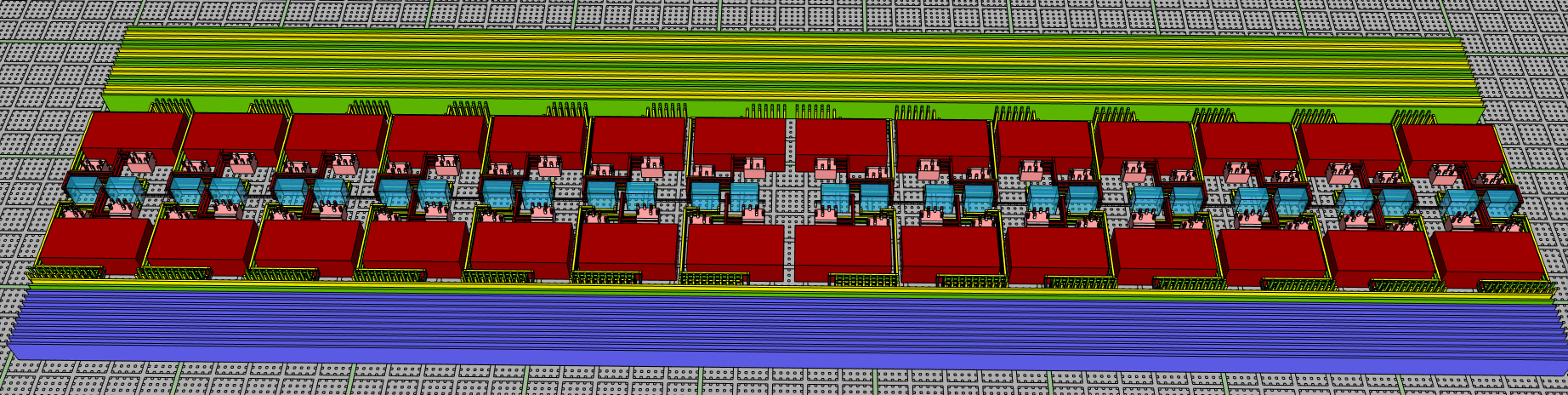}
    }
    \end{center}
    \caption{
        Spacetime layout during part of an addition.
        Time moves upward and has been truncated.
        Corresponds to one of the ``addition" columns from \fig{time-bars}.
        Tiny gray squares are the size of one logical qubit.
        Yellow and green rows are qubits from the target and input registers of the addition.
        Dark blue rows are qubits from an idle register.
        Red boxes are CCZ magic state factories, turned into AutoCCZ states by the pink boxes.
        Light blue boxes are the MAJ operation of Cuccaro's adder, arranged into a spacelike sweep to keep ahead of the control software).
        The full adder is formed by repeating this pattern, with the operating area gradually sweeping up and then down through the green/yellow data.
        The diagram is to scale for $n=2048$ with a level 2 code distance of $d=27$ and a level 1 code distance of $17$, and fits on a $225 \times 63$ rectangular grid of logical qubits.
    }
    \label{fig:addition-layout-3d}
\end{figure}

In order to perform piecewise additions separated by oblivious carry runways, we simply partition the computation horizontally as shown in \fig{addition-layout-2d} and \fig{addition-layout-3d}.
The lookup before each addition prepares registers across the pieces, as shown in \fig{lookup-layout}, but the additions themselves stick to their own piece.

The width of each piece is determined by the number of CCZ factories needed to run at the reaction limited rate.
This number is 14, assuming a code distance of 27 and using the CCZ factories from \cite{gidney2019autoccz, gidney2018magic}.
Also, assuming a level~1 code distance of 17, the footprint of the factory is~$15 \times 8$~\cite{gidney2019autoccz}.
The factories are laid out into 2 rows of 7, with single logical qubit gaps to allow space for data qubits to be routed in and out.
The total width of a piece is $15 \cdot 7 + 7 = 113$ logical qubits.
The height of the operating area is 33 rows of logical qubits ($2 \cdot 8$ for the CCZ factories, three for the ripple-carry operating area, six for AutoCCZ fixup boxes, and eight for routing data qubits).
The $\gpad + \gsep$ qubits from each register add another 30 rows (approximately).
So, overall, when working on a problem of size $n$, the computation covers a $113w \times 63$ grid of logical qubits where~$w=n/\gsep=n/1024$.

\subsection{Runtime}

Because our implementation is dominated almost entirely by the cost of performing lookup additions, its runtime is approximately equal to the number of lookup additions times the runtime of a single lookup addition.

During the lookup phase of a lookup addition, the computation is code depth limited.
Assuming a code depth of $d=27$ and a surface code cycle time of 1 microsecond, it takes $1 \mu s \cdot d/2 \cdot 2^{\gexp + \gmul} \approx 14$ milliseconds to perform the lookup using double-access hallways.

During the addition phase of a lookup addition, the computation is reaction limited.
Given a reaction time of 10 microseconds, it takes $2(\gsep + \gpad) \cdot 10 \mu s \approx 22$ milliseconds to perform the addition.
The remaining bits of a lookup addition, such as uncomputing the lookup and rearranging the rows, take approximately 1 millisecond.
Thus one lookup addition takes approximately 37~milliseconds.
Given this fact, i.e. that we perform quantum lookup additions slower than most video games render entire frames, we can approximate the total runtime:

\begin{equation}
\begin{aligned}
    \text{TotalRuntime}(n, \lenexp)
    &\approx \text{LookupAdditionCount}(n, \lenexp) \cdot 37 \;\text{milliseconds}
    \\&\approx 4 \lenexp n \;\text{milliseconds}
\end{aligned}
\end{equation}

Though we caution the reader that this estimate ignores the fact that, at larger problem sizes, lookups become slower due to the minimum code distance increasing.

This estimate implies that factoring a 2048 bit integer will take approximately 7 hours, assuming only one run of the quantum part of the algorithm is needed.
Note that in our reported numerical estimates we achieve lower per-run numbers by using more precise intermediate values and by more carefully selecting parameters.

\subsection{Distillation error}

In order to perform the $0.2 \lenexp n^2 + 0.0003 \lenexp n^2 \lg n$ Toffoli gates our implementation uses to factor an $n=2048$ bit RSA integer, we need to distill approximately 3 billion CCZ states.
According to the spreadsheet included in \cite{gidney2018magic}, using a level 1 code distance 17 and a level 2 code distance of 27, this corresponds to a total distillation error of 6.4\%.

This quantity is computed by considering the initial error rate of injecting physical T states, topological error within the factory, and the likelihood of the various stages of distillation producing a false negative.

\subsection{Topological error}

Elsewhere in this paper we frame our hardware assumptions in terms of a physical gate error rate.
In \cite{fowler2013surfaceblock} it is stated that, for a physical gate error rate of $10^{-3}$, the probability of error in a logical qubit of distance $d$, per surface code cycle, is approximately $10^{-\lceil d/2+1 \rceil}$.
We base our cost estimates on the assumption that this scaling relationship holds.
Each time the code distance is increased by two, the logical error suppression must jump by at least a factor of 10.
We believe a physical error rate of $10^{-3}$ is sufficient to achieve this scaling relationship.

Now that we know the number of logical qubits, the runtime of the algorithm, and the relationship between code distance and logical error rates, we can approximate the probability of a topological error occurring within the surface code during the execution of the algorithm.
This will allow us to verify our initial assumption that a code distance of 27 is sufficient in the case where $n=2048$.
Larger computations will require larger code distances.

When factoring an $n=2048$ bit RSA integer we are using a board of $226 \cdot 63$ logical qubits.
Approximately 25\% of these qubits are being used for distillation, which we already accounted for, and so we do not count them in this calculation.
The remaining qubits are kept through $4 \lenexp n \cdot 1000 \approx 25$ billion surface code cycles, which implies that the probability of a topological error arising is approximately $10^{-\lceil 27/2+1 \rceil} \cdot 226 \cdot 63 \cdot 0.75 \cdot 25 \cdot 10^9 \approx 27\%$.

This is a large error rate.
Using a code distance of 27 is pushing the limits of feasibility.
We would need to repeat the computation roughly 1.4 times, on average, to factor a number.
If our goal is to minimize the expected spacetime volume of the computation, perhaps we should increase the code distance to 29.
Doing so would increase the physical qubit count by 15\%, but the error rate would drop by approximately a factor of 10 and so the expected number of runs would be much closer to 1.
Ultimately the choice comes down to one's preferences for using more space versus taking more time.

\subsection{Physical qubit count}

\begin{figure}[ht]
    \begin{center}
    \resizebox{0.65 \linewidth}{!}{
        \includegraphics{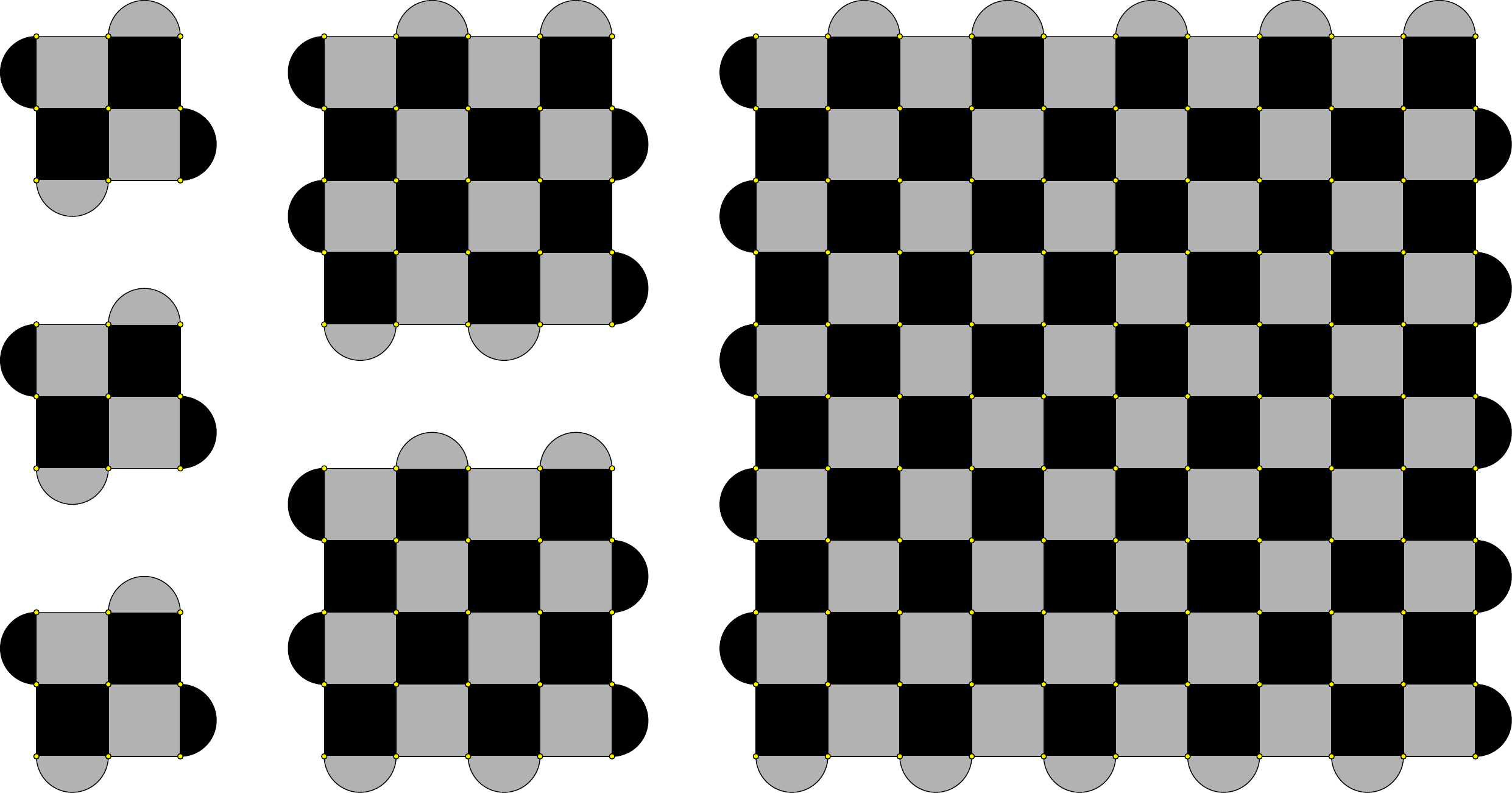}
    }
    \end{center}
    \caption{
        Stabilizer configuration diagrams for lattice surgery qubits in the rotated surface code with code distances $d$ of (from left to right) 3, 5, and 11.
        Each small yellow circle is a data qubit.
        Each filled shape is a stabilizer that is measured each surface code cycle.
        Black squares are four body Z stabilizers, gray squares are four body X stabilizers, black semi-circles are two body Z stabilizers, and gray semi-circles are two body X stabilizers.
        Not all physical qubits are shown; in particular there is a measurement qubit present for each stabilizer.
        There are $2 (d+1)^2$ physical qubits per logical qubit, including the spacing between logical qubits.
        We assume that each increase of the code distance by 2 suppresses logical errors per surface code cycle by a factor of 10 and that this is achieved with a physical error rate of $10^{-3}$.
    }
    \label{fig:lattice-surgery-qubit}
\end{figure}

In lattice surgery, a logical qubit covers $2(d+1)^2$ physical qubits where $d$ is the code distance (see \fig{lattice-surgery-qubit}).
Assuming we push the limits and use a code distance of 27 at $n=2048$, each logical qubit will cover 1568 physical qubits.
Therefore the total physical qubit count is the number of logical qubits $226 \cdot 63$ times 1568; approximately 23 million qubits.

Attentive readers will note that this number disagrees with the estimate in the title of the paper.
This is because, throughout this section, we have been sloppily rounding quantities up and choosing fixed parameters in order to keep things simple.
The estimate in the title is produced by the ancillary file ``estimate\_costs.py", which does not make these simplifications.
(In particular, ``estimate\_costs.py" realizes that the level 1 code distance used during distillation can be reduced from 17 to 15 when $n=2048$ and this adjusts the layout in several fortuitous ways.)

\subsection{Construction summary}

We now review the basic flow of our implementation of Shor's algorithm, including some details we would have otherwise left unstated.

The quantum part of the algorithm starts by preparing two registers, storing zero and one respectively, into the coset representation of modular integers with oblivious carry runways.
The cost of this preparation is negligible compared to the cost of the rest of the algorithm.
The register storing one is the accumulation register that will ultimately store the modular exponentiation, while the zero register is a workspace register.

The bulk of the execution time is then spent performing the modular exponentiation(s).
Exponent qubits are iteratively introduced in groups of size $\gexp$.
We are performing a semi-classical Fourier transform \cite{griffiths1996semiclassical}, so the exponent qubits must be phased according to measurements on previous exponent qubits.
Performing the phasing operations requires using T factories instead of CCZ factories, but there is so little phasing work compared to CCZ work that we ignore this cost in our estimates.
The phased exponent qubits are used as address qubits during a windowed modular multiplication, then measured in the frequency basis and discarded.
See \fig{exponent-grouping}.
Within this step, almost all of the computation time is spent on lookup additions.

After all of the modular multiplications have been completed, the accumulation register is storing the result of the modular exponentiation.
Note that this register still has oblivious carry runways and is still using the coset representation of modular integers.
We could decode the register before measuring it but, because the decoding operations are all classical, it is more efficient to simply measure the entire register and perform the decoding classically.
As with the cost of initialization of the registers, this cost is completely negligible.

This completes the quantum part of the algorithm.
The exponent qubit measurement results, and the decoded accumulator measurement result, are fed into the classical post-processing code which uses them to derive the solution.
If this fails, which can occur e.g. due to a distillation error during quantum execution, the algorithm is restarted.

Note that there are many minor implementation details that we have not discussed.
For example, during a windowed multiplication, one of the registers is essentially sitting idle except that its qubits are being used ($\gmul$ at a time) as address qubits for the lookup additions.
We have not discussed how these qubits are routed into and out of the lookup computation as they are needed.
It is simply clear that there is enough leeway, in the packing of the computation into spacetime, for this routing task to be feasible and contribute negligible cost.

\subsection{Other physical gate error rates}

Throughout this paper, we focus primarily on a physical gate error rate of $0.1\%$, because we consider this error rate to be plausibly achievable by quantum hardware, yet sufficiently low to enable error correction with the surface code.

This being said, it is of course important to be conservative when estimating the security of cryptographic schemes.
The reader may therefore also be interested in estimates for lower error rates.
With this in mind, in addition to including a $0.01\%$ curve in \fig{plot-costs}, we present the following rule of thumb for translating to other error rates:
$$V(p) \approx V(0.1\%) / (-\log_{10}(p) - 2)^3$$

Here $V(p)$ is the expected spacetime volume, assuming a physical gate error rate of $p$.
The rule of thumb is based on the following observation:

Each time the code distance of a construction is increased by 2, logical errors are suppressed by a factor roughly equal to the ratio between the code's threshold error rate and the hardware's physical error rate.
The surface code has a threshold error rate of around $1\%$.
For example, this means that at a physical error rate of $0.01\%$ you suppress logical errors by 100x each time you increase the code distance by 2, whereas at a physical error rate of $0.1\%$ the same suppression factor would require increasing the code distance by 4.
Consequently, code distances tend to be half as large at a physical error rate of $0.01\%$.
The code distance is a linear measure, and we are packing the computation into a three-dimensional space (one of the dimensions being time), so proportional improvements to the code distance get cubed.

Hence the idea is that at a physical error rate of $0.01\%$ code distances are half as big, meaning the computational pieces being packed into spacetime are $2^3 = 8$ times smaller, meaning in turn that the optimal packing will also be roughly eight times smaller.
Similarly, at an error rate of $0.001\%$, code distances are a third as big, so spacetime volumes should be roughly 27 times smaller.

Of course, this rule of thumb is imperfect.
For example, it does not account for the fact that code distances have to be rounded to integer values, or for the fact that the spacetime tradeoffs being made have to change due to the reaction time of the control system staying constant as the code distance changes, or for the fact that at really low error rates you would switch to different error correcting codes.
Regardless, for order of magnitude estimates involving physical error rates between $0.3\%$ and $0.001\%$, we consider this rule of thumb to be a sufficient guide.

\section{Cryptographic implications of our construction}
\label{sec:impact}
In this section, we consider how the security of RSA, and of cryptographic schemes based on the intractability of the DLP in finite fields, is affected when the optimized construction previously described is used to implement the currently most efficient derivatives of Shor's algorithms.

Our goal throughout this section is to minimize the overall expected spacetime volume, including expected repetitions, when factoring RSA integers or computing discrete logarithms. That is to say, we minimize the spacetime volume in each run of the quantum algorithm times the expected number of runs required.

\subsection{Methodology}
\label{sec:methodology}
To minimize the spacetime volume for a given choice of $n$ and $\lenexp$, we consider all combinations of
  level~1 and~2 surface code distances $\distone \in \{ 15, \, 17, \, \dots, \, 23 \}$ and $\disttwo \in \{ 25, \, 27, \, \dots, \, 51 \}$ used during distillation and computation,
  window sizes $\gmul \in \{ 4, \, 5, \, 6 \}$ and $\gexp \in \{ 4, \, 5, \, 6 \}$,
  runway spacings $\gsep \in \{ 512, \, 768, \, 1024, \, 1536, \, 2048 \}$,
  and padding offsets $\devoff \in \{ 2, \, 3, \, \dots, \, 10 \}$ where $\devoff = \gpad - 2 \lg n - \lg \lenexp$.
Furthermore, we consider two different magic state distillation strategies: the CCZ factory from \cite{gidney2018magic, gidney2019autoccz} and the T factory from \cite{fowler2018}.
For each combination of parameters we estimate the execution time $t$ and physical qubit count $s$, and upper bound the overall probability of errors occurring (to obtain the ``retry risk" $\epsilon$).

To derive an upper bound on the overall probability of errors occurring, we separately estimate the probabilities
  of topological errors occurring due to a failure of the surface code to correct errors,
  of approximation errors occurring due to using oblivious carry runways and the coset representation of modular integers,
  of magic state distillation errors,
  and of the classical post-processing algorithm failing to recover the solution from a correct run of the quantum algorithm.
We combine these error probabilities, assuming independence, to derive an upper bound on the overall probability of errors occurring.

We chose to optimize the quantity $s^{1.2} \cdot t / (1 - \epsilon)$, which we refer to as the ``skewed expected spacetime volume".
The $t / (1 - \epsilon)$ factor is the expected runtime, and the $s^{1.2}$ factor grows slightly faster than the space usage.
We skew the space usage when optimizing because we have a slight preference for decreasing space usage over decreasing runtime.
We consider all combinations of parameters $(\distone, \, \disttwo, \, \devoff, \, \gmul, \, \gexp, \, \gsep)$, choose the set that minimizes the skewed expected spacetime volume, and report the corresponding estimated costs.

\subsection{Implications for RSA}
Today, the arguably most commonly used modulus size for RSA is $n=2048$ bits.
Larger moduli are however in widespread use and smaller moduli have been used historically.

The best published academic record is the factorization of an $829$~bit RSA modulus in~2020, see~\cite{boudot-rsa829} for details regarding this computation. For the earlier record, see~\cite{kleinjung-rsa768}.

In~\tbl{impact-ifp-rsa} and \fig{plot-costs}, we provide estimates for the resource and time requirements for attacking RSA for various cryptographically relevant modulus lengths $n$. The estimates are for factoring RSA integers with Ekerå-Håstad's algorithm \cite{ekeraa2017quantum,  ekeraa2017pp} that computes a short discrete logarithm, see the appendix to \cite{ekeraa2017pp} for full technical details. As is explained in~\cite{ekeraa2017pp}, a single correct run of this quantum algorithm suffices for the RSA integer to be factored with at least $99\%$ success probability in the classical post-processing.

\begin{table}[h!]
\resizebox{0.85 \linewidth}{!}{
\begin{tabularx}{\linewidth}{
  c| 
  c|| 
  >{\centering\arraybackslash}p{5mm}| 
  >{\centering\arraybackslash}p{5mm}|| 
  >{\centering\arraybackslash}p{5mm}|| 
  >{\centering\arraybackslash}p{7mm}| 
  >{\centering\arraybackslash}p{7mm}| 
  >{\centering\arraybackslash}p{7mm}|| 
  c|| 
  >{\centering\arraybackslash}p{12.5mm}| 
  >{\centering\arraybackslash}p{12.5mm}| 
  c| 
  >{\centering\arraybackslash}p{12.5mm} 
  }
\multicolumn{2}{c}{} &
\multicolumn{6}{c}{} &
&
\multicolumn{2}{c|}{\tableheadingfont Volume} &
\tableheadingfont Qubits &
\multicolumn{1}{c}{\tableheadingfont Runtime}
\\
\multicolumn{2}{c||}{} & 
\multicolumn{6}{c||}{\tableheadingfont Parameters} &
\tableheadingfont Retry &
\multicolumn{2}{c|}{\tableheadingfont (megaqubitdays)} &
\tableheadingfont (megaqubits) &
\multicolumn{1}{c}{\tableheadingfont (hours)}
\\
$n$ &
$\lenexp$ &
$\distone$ &
$\disttwo$ &
$\devoff$ &
$\gmul$ &
$\gexp$ &
$\gsep$ &
\tableheadingfont Risk &
\tableheadingfont per run & \tableheadingfont expected &
\tableheadingfont per run &
\tableheadingfont per run
\\
\cline{1-13}
$1024$    &\multirow{7}{*}[-1pt]{\rotatebox{90}{$3(n/2-1)-40$}}
              &$15$      &$27$      &$5$       &$5$       &$5$       &$1024$    &$6\%$     &$0.5$     &$0.5$     &$9.7$     &$1.3$ \\
$2048$    &   &$15$      &$27$      &$4$       &$5$       &$5$       &$1024$    &$31\%$    &$4.1$     &$5.9$     &$20$      &$5.1$ \\
$3072$    &   &$17$      &$29$      &$6$       &$4$       &$5$       &$1024$    &$9\%$     &$19$      &$21$      &$38$      &$12$  \\
$4096$    &   &$17$      &$31$      &$9$       &$4$       &$5$       &$1024$    &$5\%$     &$48$      &$51$      &$55$      &$22$  \\
$8192$    &   &$19$      &$33$      &$4$       &$4$       &$5$       &$1024$    &$5\%$     &$480$     &$510$     &$140$     &$86$  \\
$12288$   &   &$19$      &$33$      &$3$       &$4$       &$5$       &$1024$    &$12\%$    &$1700$    &$1900$    &$200$     &$200$ \\
$16384$   &   &$19$      &$33$      &$4$       &$4$       &$5$       &$1024$    &$24\%$    &$3900$    &$5100$    &$270$     &$350$ \\
\end{tabularx}
}
\caption{Factoring an $n$ bit RSA integer by computing a short discrete logarithm. This table was produced by the script in the ancillary file ``estimate\_costs.py".}
\label{tbl:impact-ifp-rsa}
\end{table}

\subsection{Implications for finite field discrete logarithms}
Given a generator $\gen$ of an order $r$ subgroup to $\mathbb Z^*_N$, where the modulus $N$ is prime, and an element $x = \gen^d$, the finite field discrete logarithm problem is to compute $d = \log_{\gen} x$. In what follows, we assume $r$ to be prime. If $r$ is composite, the discrete logarithm problem may be decomposed into problems in subgroups of orders dividing $r$, as shown by Pohlig and Hellman \cite{pohlig-hellman}. For this reason, prime order subgroups are used in cryptographic applications.

As $\mathbb Z_N^*$ has order $N-1$, it must be that $r$ divides $N-1$, so $N = 2rk+1$ for some integer $k \ge 1$. The asymptotically best currently known classical algorithms for computing discrete logarithms in subgroups of this form are generic cycle-finding algorithms, such as Pollard's $\rho$- and $\lambda$-algorithms \cite{pollard-rho-lambda}, that run in time $O(\sqrt{r})$ and $O(\sqrt{d})$, respectively, and the general number field sieve (GNFS), that runs in subexponential time in the bit length $n$ of $N$.

The  idea of factoring via algebraic number fields was originally introduced by Pollard \cite{pollard-cubic} for integers on special forms. It was over time generalized, by amongst others Buhler et al. \cite{buhler-nfs}, Lenstra et al. \cite{lenstra-nfs} and Pollard \cite{pollard-nfs-sieving}, to factor general integers, and modified by amongst others Gordon \cite{gordon} and Schirokauer \cite{schirokauer-thesis, schirokauer} to compute discrete logarithms in finite fields. Much research effort has since been devoted to optimizing the GNFS in various respects. For an in-depth historical account of the development of the GNFS, see \cite{nfs-book, pomerance1996atale}. For the best academic record, see~\cite{boudot-rsa829}.

Let $z$ be the number of bits of security provided by the modulus with respect to classical attacks using the GNFS. Let $n_d$ and $n_r$ be the lengths in bits of $d$ and $r$, respectively. It then suffices to pick $n_d, n_r \ge 2z$ to achieve $z$ bits of classical security, as the generic cycle-finding algorithms are then not more efficient than the GNFS.

When instantiating schemes based on the intractability of the finite field discrete logarithm problem, one may hence choose between using a Schnorr group, for which $n_d = n_r = 2z$, or a safe-prime group, for which $n_r = n - 1$. In the latter case, one may in turn choose between using a short exponent, such that $n_d = 2z$, or a full length exponent, such that $n_d = n_r = n - 1$. All three parameterization options provide $z$ bits of classical security.

\subsubsection{What finite field groups are used in practice?}
In practice, Schnorr groups or safe-prime groups with short exponents are often preferred over safe-prime groups with full length exponents, as the comparatively short exponents yield considerable performance improvements.

The discrete logarithm problem in Schnorr groups is of standard form, unlike the short discrete logarithm problem in safe-prime groups, and Schnorr groups are faster to generate than safe-prime groups. A downside to using Schnorr groups is that group elements received from untrusted parties must be tested for membership of the order $r$ subgroup. This typically involves exponentiating the element to the power of $r$, which is computationally expensive. Safe-prime groups are more flexible than Schnorr groups, in that the exponent length may be adaptively selected depending on the performance requirements. The reader is referred to \cite{oorschot} for a more in-depth comparison and historical recommendations. In more recent years, the use of safe-prime groups would appear to have become increasingly prevalent. Some cryptographic schemes, such as the Diffie-Hellman key agreement protocol, are agnostic to the choice of group, whereas other schemes, such as DSA, use Schnorr groups for efficiency reasons.

The National Institute of Standards and Technology (NIST) in the United States standardizes the use of crypto\-graphy in unclassified applications within the federal government. Up until April of 2018, NIST recommended the use of randomly selected Schnorr groups with moduli of length 2048 bits for Diffie-Hellman key agreement. NIST changed this recommendation in the 3rd revision of SP800-56A \cite{nist-sp-800-56-part1-rev3-2018}, and are now advocating using a fixed set of safe-prime groups with moduli of length up to 8192 bits, with short or full length exponents. These groups were originally developed for TLS \cite{rfc-tls} and IKE \cite{rfc-ike} where, again, they are used either with short of full length exponents.

\subsubsection{Complexity estimates}
To estimates the resource and time requirements for computing discrete logarithms in finite fields for various modulus lengths $n$, and for the aforementioned parameterization options, we need to decide on what model to use for estimating $z$ as a function of $n$. Various models have been proposed over the years, see for instance \cite{lenstra-verheul-model-2001, lenstra-model-2004, keylength2019}. For simplicity, we use the same model that NIST uses in SP 800-56A \cite{nist-sp-800-56-part1-rev3-2018}. It is described on p.~110 of FIPS 140-2 IG \cite{fips-140-2-IG}. Note that NIST rounds $z$ to the closest multiple of eight bits.

To compute short logarithms in safe-prime groups, the best option is to use Ekerå-Håstad's algorithm \cite{ekeraa2017quantum, ekeraa2017pp} that is specialized for this purpose. To compute general discrete logarithms in safe-prime or Schnorr groups, one option is to use Ekerå's algorithm \cite{ekeraa2018general}. As is explained in~\cite{ekeraa2017pp, ekeraa2018general}, a single correct run of these quantum algorithms suffices for the logarithm to be recovered with~$\ge 99\%$ success probability in the classical post-processing. These algorithms do not require the order of the group to be known. See~\tbl{impact-dlp-ff} and \fig{plot-costs} for complexity estimates.

\begin{table}[h!]
\resizebox{0.69 \linewidth}{!}{
\begin{tabularx}{\linewidth}{
  >{\centering\arraybackslash}p{4mm} 
  >{\centering\arraybackslash}p{4mm} 
  c| 
  >{\centering\arraybackslash}p{4mm}|| 
  >{\centering\arraybackslash}p{4mm}| 
  >{\centering\arraybackslash}p{4mm}| 
  c|| 
  >{\centering\arraybackslash}p{5mm}| 
  >{\centering\arraybackslash}p{5mm}|| 
  >{\centering\arraybackslash}p{5mm}|| 
  >{\centering\arraybackslash}p{7mm}| 
  >{\centering\arraybackslash}p{7mm}| 
  >{\centering\arraybackslash}p{7mm}|| 
  c|| 
  >{\centering\arraybackslash}p{12.5mm}| 
  >{\centering\arraybackslash}p{12.5mm}| 
  c| 
  >{\centering\arraybackslash}p{12.5mm} 
  }
&& 
\multicolumn{2}{c}{} & 
\multicolumn{3}{c}{} & 
\multicolumn{6}{c}{} &
&
\multicolumn{2}{c|}{\tableheadingfont Volume} &
Qubits &
\multicolumn{1}{c}{\tableheadingfont Runtime}
\\
&& 
\multicolumn{2}{c}{} & 
\multicolumn{3}{c||}{} & 
\multicolumn{6}{c||}{\tableheadingfont Parameters} &
\tableheadingfont Retry &
\multicolumn{2}{c|}{\tableheadingfont (megaqubitdays)} &
\tableheadingfont (megaqubits) &
\multicolumn{1}{c}{\tableheadingfont (hours)}
\\
&& 
$n$ &
$\lenexp$ &
$n_d$ &
$n_r$ &
$z$ &
$\distone$ &
$\disttwo$ &
$\devoff$ &
$\gmul$ &
$\gexp$ &
$\gsep$ &
\tableheadingfont  Risk &
\tableheadingfont per run & \tableheadingfont expected &
\tableheadingfont per run &
\tableheadingfont per run
\\
\cline{1-18}
\multirow{7}{*}[0pt]{\rotatebox{90}{\tableheadingfont Schnorr}} &
  & $1024$    &\multirow{7}{*}[0pt]{\rotatebox{90}{$3 n_r = 6z$}}     & \multirow{7}{*}[0pt]{$2z$} & \multirow{7}{*}[0pt]{$2z$} & $80$   &$15$      &$25$      &$4$       &$5$       &$5$       &$1024$    &$10\%$    &$0.2$     &$0.2$     &$9.2$     &$0.4$ \\
& & $2048$    &    & & & $112$  &$15$      &$27$      &$3$       &$5$       &$5$       &$1024$    &$9\%$     &$0.9$     &$1.0$     &$20$      &$1.2$ \\
& & $3072$    &    & & & $128$  &$15$      &$27$      &$9$       &$5$       &$5$       &$1024$    &$18\%$    &$2.4$     &$2.9$     &$29$      &$2.0$ \\
& & $4096$    &    & & & $152$  &$17$      &$29$      &$7$       &$4$       &$5$       &$1024$    &$4\%$     &$6.5$     &$6.8$     &$51$      &$3.1$ \\
& & $8192$    &    & & & $200$  &$17$      &$31$      &$3$       &$4$       &$5$       &$1024$    &$5\%$     &$38$      &$40$      &$110$     &$8.3$ \\
& & $12288$   &    & & & $240$  &$17$      &$31$      &$9$       &$4$       &$5$       &$1024$    &$9\%$     &$110$     &$120$     &$170$     &$15$  \\
& & $16384$   &    & & & $272$  &$17$      &$31$      &$5$       &$4$       &$5$       &$1024$    &$17\%$    &$210$     &$250$     &$220$     &$23$  \\
\cline{1-18}
\multirow{14}{*}[0pt]{\rotatebox{90}{\tableheadingfont Safe-prime}} &
\multirow{7}{*}[0pt]{\rotatebox{90}{\tableheadingfont Short}} &
  $1024$      &\multirow{7}{*}[0pt]{\rotatebox{90}{$3 n_d = 6z$}}     & \multirow{7}{*}[0pt]{$2z$} & \multirow{7}{*}[0pt]{\rotatebox{90}{$n-1$}}  & $80$   &$15$      &$25$      &$4$       &$5$       &$5$       &$1024$    &$10\%$    &$0.2$     &$0.2$     &$9.2$     &$0.4$ \\
& & $2048$    &    & & & $112$  &$15$      &$27$      &$3$       &$5$       &$5$       &$1024$    &$9\%$     &$0.9$     &$1.0$     &$20$      &$1.2$ \\
& & $3072$    &    & & & $128$  &$15$      &$27$      &$9$       &$5$       &$5$       &$1024$    &$18\%$    &$2.4$     &$2.9$     &$29$      &$2.0$ \\
& & $4096$    &    & & & $152$  &$17$      &$29$      &$7$       &$4$       &$5$       &$1024$    &$4\%$     &$6.5$     &$6.8$     &$51$      &$3.1$ \\
& & $8192$    &    & & & $200$  &$17$      &$31$      &$3$       &$4$       &$5$       &$1024$    &$5\%$     &$38$      &$40$      &$110$     &$8.3$ \\
& & $12288$   &    & & & $240$  &$17$      &$31$      &$9$       &$4$       &$5$       &$1024$    &$9\%$     &$110$     &$120$     &$170$     &$15$  \\
& & $16384$   &    & & & $272$  &$17$      &$31$      &$5$       &$4$       &$5$       &$1024$    &$17\%$    &$210$     &$250$     &$220$     &$23$  \\
\cline{2-18}
& \multirow{7}{*}[0pt]{\rotatebox{90}{\tableheadingfont Full Length}} &
    $1024$    &\multirow{7}{*}[0pt]{\rotatebox{90}{$3n_r = 3(n-1)$}}     & \multirow{7}{*}[0pt]{\rotatebox{90}{$n-1$}}  & \multirow{7}{*}[0pt]{\rotatebox{90}{$n-1$}}  & $80$     &$15$      &$27$      &$9$       &$5$       &$5$       &$1024$    &$10\%$    &$1.1$     &$1.2$     &$9.7$     &$2.7$ \\
& & $2048$    & & & & $112$    &$17$      &$29$      &$6$       &$4$       &$5$       &$1024$    &$6\%$     &$12$      &$12$      &$26$      &$11$  \\
& & $3072$    & & & & $128$    &$17$      &$31$      &$5$       &$4$       &$5$       &$1024$    &$5\%$     &$41$      &$43$      &$41$      &$24$  \\
& & $4096$    & & & & $152$    &$17$      &$31$      &$7$       &$4$       &$5$       &$1024$    &$9\%$     &$97$      &$110$     &$55$      &$43$  \\
& & $8192$    & & & & $200$    &$19$      &$33$      &$4$       &$4$       &$5$       &$1024$    &$8\%$     &$960$     &$1100$    &$140$     &$180$ \\
& & $12288$   & & & & $240$    &$19$      &$33$      &$3$       &$4$       &$5$       &$1024$    &$21\%$    &$3300$    &$4100$    &$200$     &$390$ \\
& & $16384$   & & & & $272$    &$21$      &$35$      &$4$       &$4$       &$5$       &$1024$    &$16\%$    &$9100$    &$11000$   &$320$     &$700$ \\
\end{tabularx}
}
\caption{Computing discrete logarithms using Ekerå-Håstad's \cite{ekeraa2017quantum, ekeraa2017pp} and Ekerå's \cite{ekeraa2018general} algorithms.
This table was produced by the script in the ancillary file ``estimate\_costs.py".}
\label{tbl:impact-dlp-ff}
\end{table}

If the group order is known, a better option for computing general discrete logarithms in safe-prime groups and Schnorr groups when not making tradeoffs is to use Shor's original algorithm~\cite{shor1994}, modified to work in the order $r$ subgroup rather than in the whole multiplicative group $\mathbb Z_N^*$, and to start with a uniform superposition of all exponent values, as opposed to superpositions of $r$ values.
Note that the latter modification is necessary to enable the use of the semi-classical Fourier transform, qubit recycling and the windowing technique.
The modified algorithm is described in~\cite{ekeraa2019revisiting} where a heuristic analysis is also provided.

\begin{table}[h!]
\resizebox{0.69 \linewidth}{!}{
\begin{tabularx}{\linewidth}{
  >{\centering\arraybackslash}p{4mm} 
  c| 
  >{\centering\arraybackslash}p{4mm}|| 
  >{\centering\arraybackslash}p{4mm}| 
  >{\centering\arraybackslash}p{4mm}| 
  c|| 
  >{\centering\arraybackslash}p{5mm}| 
  >{\centering\arraybackslash}p{5mm}|| 
  >{\centering\arraybackslash}p{5mm}|| 
  >{\centering\arraybackslash}p{7mm}| 
  >{\centering\arraybackslash}p{7mm}| 
  >{\centering\arraybackslash}p{7mm}|| 
  c|| 
  >{\centering\arraybackslash}p{12.5mm}| 
  >{\centering\arraybackslash}p{12.5mm}| 
  c| 
  >{\centering\arraybackslash}p{12.5mm} 
  }
& 
\multicolumn{2}{c}{} & 
\multicolumn{3}{c}{} & 
\multicolumn{6}{c}{} &
&
\multicolumn{2}{c|}{\tableheadingfont Volume} &
\tableheadingfont Qubits &
\multicolumn{1}{c}{\tableheadingfont Runtime}
\\
& 
\multicolumn{2}{c}{} & 
\multicolumn{3}{c||}{} & 
\multicolumn{6}{c||}{\tableheadingfont Parameters} &
\tableheadingfont Retry &
\multicolumn{2}{c|}{\tableheadingfont (megaqubitdays)} &
\tableheadingfont (megaqubits) &
\multicolumn{1}{c}{\tableheadingfont (hours)}
\\
& 
$n$ &
$\lenexp$ &
$n_d$ &
$n_r$ &
$z$ &
$\distone$ &
$\disttwo$ &
$\devoff$ &
$\gmul$ &
$\gexp$ &
$\gsep$ &
\tableheadingfont Risk &
\tableheadingfont per run & \tableheadingfont expected &
\tableheadingfont per run &
\tableheadingfont per run
\\
\cline{1-17}
\multirow{7}{*}[0pt]{\rotatebox{90}{\tableheadingfont Schnorr}} &
  $1024$    &\multirow{7}{*}[0pt]{\rotatebox{90}{$2(n_r+5)$}}     & \multirow{7}{*}[0pt]{$2z$} & \multirow{7}{*}[0pt]{$2z$} &
    $80$   &$15$      &$25$      &$6$       &$5$       &$5$       &$1024$    &$8\%$     &$0.1$     &$0.1$     &$9.2$     &$0.3$ \\
& $2048$    &    & & & $112$  &$15$      &$27$      &$8$       &$5$       &$5$       &$1024$    &$6\%$     &$0.6$     &$0.7$     &$20$      &$0.8$ \\
& $3072$    &    & & & $128$  &$15$      &$27$      &$6$       &$5$       &$5$       &$1024$    &$13\%$    &$1.6$     &$1.8$     &$29$      &$1.3$ \\
& $4096$    &    & & & $152$  &$15$      &$27$      &$3$       &$5$       &$5$       &$1024$    &$25\%$    &$3.3$     &$4.4$     &$39$      &$2.1$ \\
& $8192$    &    & & & $200$  &$17$      &$29$      &$5$       &$4$       &$5$       &$1024$    &$11\%$    &$23$      &$26$      &$110$     &$5.5$ \\
& $12288$   &    & & & $240$  &$17$      &$31$      &$4$       &$4$       &$5$       &$1024$    &$8\%$     &$68$      &$74$      &$170$     &$10$  \\
& $16384$   &    & & & $272$  &$17$      &$31$      &$5$       &$4$       &$5$       &$1024$    &$12\%$    &$140$     &$160$     &$220$     &$16$  \\
\cline{1-17}
\multirow{7}{*}[0pt]{\rotatebox{90}{\tableheadingfont Safe-prime}} &
  $1024$    &\multirow{7}{*}[0pt]{\rotatebox{90}{$2(n_r+5)$}}     & \multirow{7}{*}[0pt]{\rotatebox{90}{$n-1$}}  & \multirow{7}{*}[0pt]{\rotatebox{90}{$n-1$}}  &
    $80$     &$15$      &$27$      &$3$       &$5$       &$5$       &$1024$    &$8\%$     &$0.7$     &$0.8$     &$9.7$     &$1.8$ \\
& $2048$    & & & & $112$    &$17$      &$29$      &$3$       &$4$       &$5$       &$1024$    &$5\%$     &$7.4$     &$7.8$     &$26$      &$7.0$ \\
& $3072$    & & & & $128$    &$17$      &$29$      &$4$       &$4$       &$5$       &$1024$    &$12\%$    &$25$      &$29$      &$38$      &$16$  \\
& $4096$    & & & & $152$    &$17$      &$31$      &$4$       &$4$       &$5$       &$1024$    &$8\%$     &$65$      &$70$      &$55$      &$29$  \\
& $8192$    & & & & $200$    &$19$      &$33$      &$3$       &$4$       &$5$       &$1024$    &$6\%$     &$640$     &$680$     &$140$     &$120$ \\
& $12288$   & & & & $240$    &$19$      &$33$      &$4$       &$4$       &$5$       &$1024$    &$15\%$    &$2200$    &$2600$    &$200$     &$260$ \\
& $16384$   & & & & $272$    &$21$      &$35$      &$2$       &$4$       &$5$       &$1024$    &$12\%$    &$6100$    &$6900$    &$320$     &$470$ \\
\end{tabularx}
}
\caption{Computing discrete logarithms using Shor's algorithm \cite{shor1994} modified as described in \cite{ekeraa2016modifying, ekeraa2019revisiting}.
This table was produced by the script in the ancillary file ``estimate\_costs.py".}
\label{tbl:impact-dlp-ff-shor}
\end{table}

When using this modified version of Shor's algorithm to compute discrete logarithms, the heuristic analysis \cite{ekeraa2019revisiting} shows that a single correct run suffices to compute the logarithm with~$\ge 99\%$ success probability, assuming that each of the two exponent registers is padded with $5$ bits, and that a small search is performed in the classical post-processing.
This implies that Shor's algorithm outperforms Ekerå's algorithm, as Shor's algorithm performs only approximately $2n_r$ group operations per run, compared to $3n_r$ operations in Ekerå's algorithm, see \tbl{impact-dlp-ff-shor} for complexity estimates. This is because Ekerå's algorithm does not require $r$ to be known. In fact, it computes both $d$ and $r$.

Note that for safe-prime groups, $r = (N-1)/2$, so when $N$ is known to the adversary then so is $r$. For Schnorr groups, it may be that $r$ is unknown to the adversary, especially if the group is randomly selected. It may be hard to compute $r$ classically, as it amounts to finding a $n_r = 2z$ bit prime factor of $(N-1)/2$.

\subsection{Implications for elliptic curve discrete logarithms}
Over the past decades cryptographic schemes based on the intractability of the DLP in finite fields and the RSA integer factoring problem have gradually been replaced by cryptography based on the intractability of the DLP in elliptic curve groups. This is reflected in standards issued by organizations such as NIST.

Not all optimizations developed in this paper are directly applicable to arithmetic in elliptic curve groups. It is an interesting topic for future research to study to what extent the optimizations developed in this paper may be adapted to optimize such arithmetic operations (see \sec{revisit-elliptic-curves}). This paper should not be perceived to indicate that the RSA integer factoring problem and the DLP in finite fields is in itself less complex than the DLP in elliptic curve groups on quantum computers. The feasibility of optimizing the latter problem must first be properly studied.

\subsection{On the importance of complexity estimates}
It is important to estimate the complexity of attacking widely deployed asymmetric cryptographic schemes using future large-scale quantum computers. Such estimates enable informed decisions to be made on when to mandate migration from existing schemes to post-quantum secure schemes.

For cryptographic schemes that are used to protect confidentiality, such as encryption and key agreement schemes, a sufficiently long period must elapse inbetween the point in time when the scheme ceases to be used, and the point in time when the scheme is projected to become susceptible to practical attacks. This is necessary so as to ensure that the information that has been afforded protection with the scheme is no longer sensitive once the scheme becomes susceptible to practical attacks. This is because one must assume that encrypted information may be recorded and archived for decryption in the future.

If the information you seek to protect is to remain confidential for 25 years, you must hence stop using asymmetric schemes such as RSA and Diffie-Hellman at least 25 years before quantum computers capable of breaking these schemes become available to the adversary. For cryptographic schemes that are used to protect authenticity, or for authentication, such as signature schemes, it suffices to migrate to post-quantum secure schemes only right before the schemes become susceptible to practical attacks. This is an important distinction.

\subsection{On early adoption of post-quantum secure schemes}
The process of transitioning to post-quantum secure schemes has already begun.
However, no established or universally recognized standards are as of yet available.
Early adopters may therefore wish to consider implementing schemes conjectured to be post-quantum secure  alongside existing classically secure schemes, in such a fashion that both schemes must be broken for the combined hybrid scheme to be broken.

\section{Future work}
\label{sec:future-work}

\subsection{Investigate asymptotically efficient multiplication}

The multiplication circuits that we are using have a Toffoli count that scales quadratically (up to polylog factors).
There are multiplication circuits with asymptotically better Toffoli counts.
For example, the Karatsuba algorithm~\cite{karatsuba1962multiplication} has a Toffoli count of $O(n^{\lg 3})$ and the Schönhage–Strassen algorithm \cite{schonhage1971multiply} has a Toffoli count of $O(n \lg n \lg \lg n)$.
However, there are difficulties when attempting to use these asymptotically efficient algorithms in the context of Shor's algorithm.

The first difficulty is that efficient multiplication algorithms are typically classical, implemented with non-reversible computation in mind.
They need to be translated into a reversible form.
This is not trivial.
For example, a naive translation of Karatsuba multiplication will result in twice as many recursive calls at each level (due to the need to uncompute), and increase the asymptotic Toffoli count from $O(n^{\lg 3})$ to $O(n^{\lg 6})$.
Attempting to fix this problem can result in the space complexity increasing \cite{parent2017karatsuba}, though it is possible to solve this problem \cite{gidney2019karatsuba}.

The second difficulty is constant factors, both in workspace and in Toffoli count.
Clever multiplication circuits have better Toffoli counts for sufficiently large $n$ but, when we do back-of-the-envelope estimates, ``sufficiently large" is beyond $n=2048$.
This difficulty is made worse if the multiplier is incompatible with optimizations that work on naive multipliers, such as windowed arithmetic and the coset representation of modular integers.
Clever multiplication circuits also tend to use additional workspace, and it is necessary to contrast using the better multiplier against the opportunity cost of using the space for other purposes (such as distillation of magic states).
For example, the first step of the Schönhage–Strassen algorithm is to pad the input up to double length, then split the padded register into $O(\sqrt{n})$ pieces and pad each piece up to double length.
The target register quadruples in size before even getting into the details of performing the number theoretic transform!
This large increase in space means that the quantum variant of the Schönhage–Strassen algorithm is competing with the many alternative opportunities one has when given 6000 more logical qubits of workspace (at $n=2048$).
For example, that's enough space for fifty additional CCZ factories.

Can multipliers with asymptotically lower Toffolis counts help at practical sizes, such as $n=2048$~bits?
We believe that they do not, but only a careful investigation can properly determine the answer to this question.

\subsection{Optimize distillation}

To produce our magic states, we use slightly-modified copies of the CCZ factory from \cite{gidney2018magic} as explained in \cite{gidney2019autoccz}.
We have many ideas for improving on this approach.

First, the factory we are using is optimized for the case where it is working in isolation.
But, generally speaking, error correcting codes get better when working over many objects instead of one object.
It is likely that a factory using a block code to produce multiple CCZ states at once would perform better than the factory we used.
For example, \cite{haah2018codes} presents a distillation protocol that produces good CCZ states ten at a time.

Second, since publishing \cite{gidney2018magic}, we have realized there are two obvious-in-hindsight techniques that could be used to reduce the volume of the factories in that paper.
First, we assumed that topological errors that can occur within the factories were undetected, but actually many of them are heralded as distillation failures.
By taking advantage of this heralding, it should be possible to reduce the code distance used in many parts of the factories.
Second, when considering the set of S gates to apply to correct T gate teleportations performed by the factories, there are redundancies that we were not previously aware of.
In particular, for each measured X stabilizer, one can toggle whether or not every qubit in that stabilizer has an S gate applied to it.
This freedom makes it possible to e.g. apply dynamically chosen corrections while guaranteeing that the number of S gate fixups in the 15-to-1 T factory is at most 5 (instead of 15), or to ensure a particular qubit will never need an S gate fixup (removing some packing constraints).

Third, we believe it should be possible to use detection events produced by the surface code to estimate how likely it is that an error occurred, and that this information can be use to discard ``risky factory runs" in a way that increases reliability.
This would allow us to trade the increased reliability for a decreased code distance.
As an extreme example, suppose that instead of attempting to correct errors during a factory run we simply discarded the run if there were any detection events where a local stabilizer flipped.
Then the probability that a factory run would complete without being discarded would be approximately zero, but when a run did pass the chance of error would be amazingly low.
We believe that by picking the right metric (e.g. number of detections or diameter of alternating tree during matching), then interpolating a rejection threshold between the extreme no-detections-anywhere rule and the implicit hope-all-errors-were-corrected rule that is used today, there will be a middle ground with lower expected volume per magic state.
(Even more promisingly, this thresholded error estimation technique should work on any small state production task and almost all quantum computation can be reformulated as a series of small state production tasks.)

Reducing the volume of distillation would improve the space and time estimates in this paper.
But turning some combination of the above ideas into a concrete factory layout, with understood error behavior, is too large of a task for one paper.
Of course, the problem of finding small factories is a problem that the field has been exploring for some time.
In fact, in the week before we first released this paper, Daniel Litinsky made \cite{litinski2019magic} available.
He independently arrived at the idea of using distillation to herald errors within the factory, and provided rough numerics indicating this may reduce the volume of distillation by as much as a factor of 10.
Therefore we leave optimizing distillation not as future work, but as ongoing work.

\subsection{Optimize qubit encoding}

\begin{figure}
    \resizebox{1.0 \linewidth}{!}{
        \includegraphics{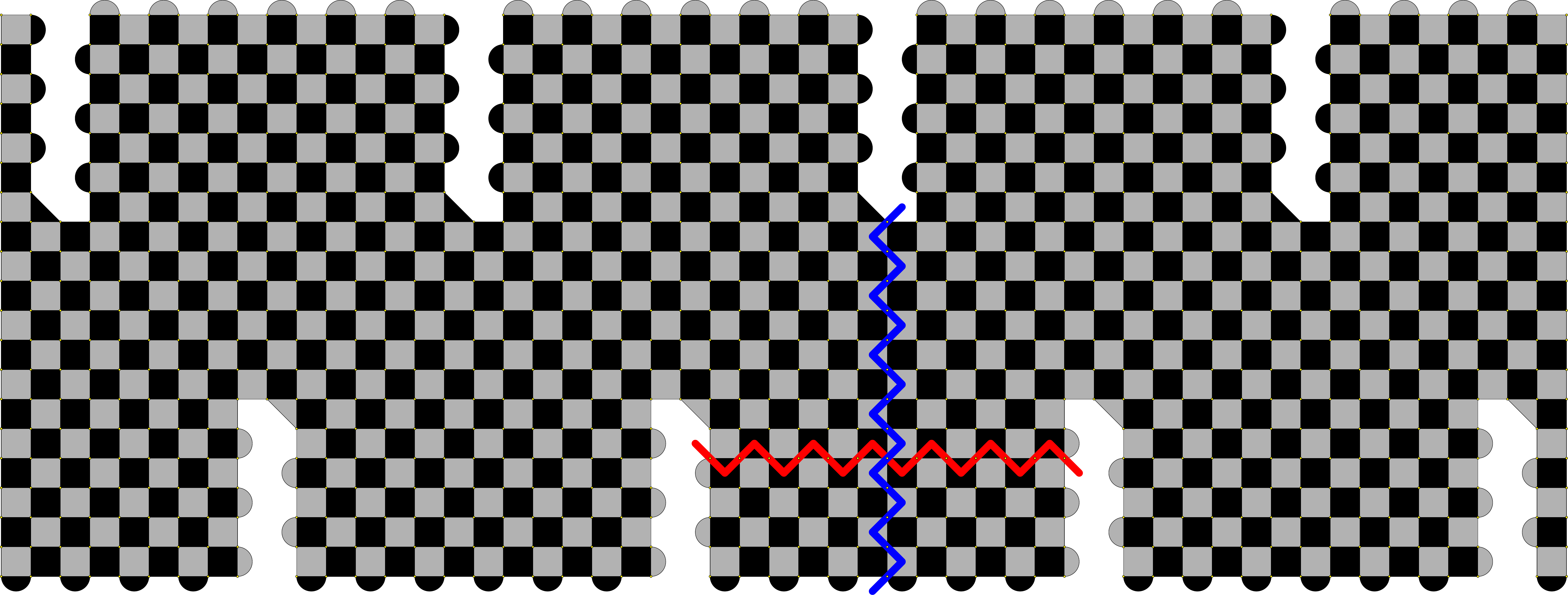}
    }
    \caption{
        A possible dense packing for idle qubits, using $1.5d^2 + O(d)$ physical qubits instead of $2(d+1)^2$.
        In the diagram, there is one distance 13 logical qubit per ``bulb".
        The X and Z observables of one of the logical qubits are shown in red and blue respectively.
        The pattern continues past the cut-off at the left and right sides.
    }
    \label{fig:qubit-houses}
\end{figure}

Most of the logical qubits in our construction spend most of their time sitting still, waiting for other qubits to be processed.
It should be possible to store these qubits in a more compact, but less computationally convenient, form.
A simple example is that resting qubits do not need the ``padding layer" between qubits implicit in \fig{lattice-surgery-qubit}, because this layer is only there to make surgery easier.
Another example is that there may to be ways to pack lattice surgery qubits that use less area while preserving the code distance (e.g. the ``bulbs" shown in \fig{qubit-houses}).

One could also imagine encoding the resting qubits into a block code.
The difficulty is in finding a block code that a) works with small groups of qubits, b) can be encoded and decoded fast enough and compactly enough to fit into the existing computation, and c) is sufficiently better than the surface code that the benefits (reduced code distance within the surface code) outweigh the costs (redundant additional qubits).

Finally, there are completely different ways of storing information in the surface code.
For example, qubits stored using dislocations \cite{hastings2014dislocations} could be denser than qubits stored using lattice surgery.
However, dislocations also create runs of stabilizers over not-quite-adjacent physical qubits.
Measuring these stabilizers requires additional physical operations, creating more opportunities for errors, and so errors will propagate more quickly along these runs.

Do the ``qubit houses" in \fig{qubit-houses} actually work, or is there some unexpected error mechanism?
Can dislocation qubits be packed more tightly than lattice surgery qubits while achieving an equivalent logical error rate?
Is there a block code that is sufficiently beneficial when layered over surface code qubits?
Until careful simulations are done it will be unclear what the answer to these questions is, and so we leave the answers to future work.

\subsection{Distribute the computation}

An interesting consequence of how we parallelize addition, by segmenting registers into pieces terminated by carry runways, is that there is limited interaction between the different pieces.
In fact, the additions themselves require {\em no} communication between the pieces; it is only the lookup procedure preparing the input of the additions that requires communication.
If we place each piece on a separate quantum computer, only a small amount of communication is needed to seed the lookups and keep the computation progressing.

Given the parameters we chose for our construction, each lookup is controlled by ten address qubits.
Five of those qubits come from the exponent and are re-used hundreds of times.
The other five come from the factor register, which is being iterated over during multiplication.
If the computation was to be distributed over multiple machines, the communication cost would be dominated by broadcasting the successive groups of five qubits from the factor register.

Recall from \sec{construction} that each lookup addition takes approximately 37 milliseconds.
Five qubits must be broadcast per lookup addition.
Therefore, if each quantum computer had a 150~qb/s quantum channel, the necessary qubits could be communicated quickly enough to keep up with the lookup additions.
This would distribute the factoring computation.

For example, a network topology where the computers were arranged into a linear chain and were constantly forwarding the next set of factor qubits to each other would be sufficient.
Each quantum computer in the distributed computation would have a computational layout basically equivalent to the left  half (or right half) of \fig{addition-layout-3d}.
Factoring an $n=2048$ bit number could be performed with two machines each having perhaps 11 million physical qubits, instead of one machine with 20 million physical qubits.

We can decrease the size of each individual quantum computer by decreasing the piece size we use for the additions.
For example, suppose we used an addition piece size of $\gsep=256$ and distributed an $n=2048$ RSA factoring computation over 8 machines.
Normally using smaller pieces would proportionally accelerate the addition, but the number of magic state factories has not changed (they have simply been distributed) so each lookup addition will still take approximately the same amount of time.
So a 150 qb/s quantum channel per computer should still be sufficient.
One caveat here is that each of these much smaller machines has to run its own copy of the lookup operation, and because there are so few factories per machine the lookup will take much longer.
The optimal windows sizes of the lookups will decrease, and the total computation time will approximately double.

Based on this surface analysis it seems that, instead of using 1 machine with 20 million qubits, we could use 8 machines each with perhaps 4 million qubits, as long as they were connected by quantum channels with bandwidths of 150qb/s.
But we have not carefully explored the question of how to distribute a quantum computation structured in this way and, although there has been significant previous work done on distributing quantum computations such as \cite{meter2008arithmetic,oi2006scalable,jiang2010scalable}, we think the piecewise nature of carry runway adders makes them well suited to a specialized analysis.

\subsection{Revisit elliptic curves}
\label{sec:revisit-elliptic-curves}

Many of the optimization techniques that we use in this paper generalize to other contexts where arithmetic is performed.
In particular, consider the Toffoli count for computing discrete logarithms over elliptic curves reported in~\cite{roetteler2017quantum}.
It can likely be improved substantially by using windowed arithmetic.
On the other hand, because of the need to compute modular inverses, it is not clear if the coset representation of modular integers is applicable.

Which of our optimizations can be ported over, and which ones cannot?
How much of an improvement would result?
These are interesting questions for future research work.

\section{Conclusion}
\label{sec:conclusion}

In this paper, we combined several techniques and optimizations into an efficient construction for factoring integers and computing discrete logarithms over finite fields.
We estimated the approximate cost of our construction, both in the abstract circuit model and under plausible physical assumptions for large-scale quantum computers based on superconducting qubits.
We presented concrete cost estimates for several cryptographically relevant problems.
Our estimated costs are orders of magnitude lower than in previous works with comparable physical assumptions.

In \cite{mosca2018cybersecurity}, Mosca poses the rhetorical question: ``How many physical qubits will we need to break RSA-2048? [...] Current estimates range from tens of millions to a billion physical qubits".
One of the sources estimating a billion physical qubits is \cite{fowler2012surfacecodereview}.
Our physical assumptions are more pessimistic than the physical assumptions used in that paper (see \tbl{historical-comparison}) so it is reasonable to say that, in the four years since 2015, the upper end of the estimate of how many qubits will be needed to factor 2048 bit RSA integers has dropped nearly two orders of magnitude; from a billion to twenty million.

Clearly the low end of Mosca's estimate should also drop.
However, the low end of the estimate is highly sensitive to advances in the design of quantum error correcting codes, the engineering of physical qubits, and the construction of quantum circuits.
Predicting such advances is beyond the scope of this paper.

Post-quantum cryptosystems are in the process of being standardized \cite{alagic2019status}, and small-scale experiments with deploying such systems on the internet have been performed \cite{google2016postquantum}.
However, a considerable amount of work remains to be done to enable large-scale deployment of post-quantum cryptosystems.
We hope that this paper informs the rate at which this work needs to proceed.

\section*{Contributions}
Craig Gidney designed the efficient construction for modular exponentiation, produced initial cost estimates for RSA, and assembled results from other papers for comparison.
Martin Ekerå did the cryptographic impact analysis, and extended the construction and cost estimates to problems beyond factoring RSA integers and to algorithms beyond Shor's factoring algorithm.

\section*{Acknowledgements}
We thank Adam Langley, Ilya Mironov, Ananth Raghunathan, and Ryan Babbush for reading a draft of this paper and providing useful feedback which improved it.
We thank Austin Fowler and Johan Håstad for useful feedback and discussions.
Craig Gidney thanks Hartmut Neven for creating an environment where this research was possible in the first place.

\bibliography{refs}

\begin{thebibliography}{91}
\providecommand{\natexlab}[1]{#1}
\providecommand{\url}[1]{\texttt{#1}}
\expandafter\ifx\csname urlstyle\endcsname\relax
  \providecommand{\doi}[1]{doi: #1}\else
  \providecommand{\doi}{doi: \begingroup \urlstyle{rm}\Url}\fi

\bibitem[Alagic et~al.(2019)Alagic, Alperin-Sheriff, Apon, Cooper, Dang, Liu,
  Miller, Moody, Peralta, Perlner, Robinson, and Smith-Tone]{alagic2019status}
G.~Alagic, J.~Alperin-Sheriff, D.~Apon, D.~Cooper, Q.~Dang, Y.-K. Liu,
  C.~Miller, D.~Moody, R.~Peralta, R.~Perlner, A.~Robinson, and D.~Smith-Tone.
\newblock {Status Report on the First Round of the NIST Post-Quantum
  Cryptography Standardization Process}.
\newblock Technical Report NIST Internal Report (NISTIR) 8240, NIST, January
  2019.
\newblock \doi{10.6028/NIST.IR.8240}.

\bibitem[Babbush et~al.(2018)Babbush, Gidney, Berry, Wiebe, McClean, Paler,
  Fowler, and Neven]{babbush2018}
R.~Babbush, C.~Gidney, D.~W. Berry, N.~Wiebe, J.~McClean, A.~Paler, A.~Fowler,
  and H.~Neven.
\newblock {Encoding Electronic Spectra in Quantum Circuits with Linear T
  Complexity}.
\newblock \emph{{Physical Review X}}, 8\penalty0 (4):\penalty0 041015(1--36),
  2018.
\newblock \doi{10.1103/PhysRevX.8.041015}.
\newblock arXiv:1805.03662.

\bibitem[Barends et~al.(2014)Barends, Kelly, Megrant, Veitia, Sank, Jeffrey,
  White, Mutus, Fowler, Campbell, Chen, Chen, Chiaro, Dunsworth, Neill,
  O'Malley, Roushan, Vainsencher, Wenner, Korotkov, Cleland, and
  Martinis]{Bare13}
R.~Barends, J.~Kelly, A.~Megrant, A.~Veitia, D.~Sank, E.~Jeffrey, T.~C. White,
  J.~Mutus, A.~G. Fowler, B.~Campbell, Y.~Chen, Z.~Chen, B.~Chiaro,
  A.~Dunsworth, C.~Neill, P.~O'Malley, P.~Roushan, A.~Vainsencher, J.~Wenner,
  A.~N. Korotkov, A.~N. Cleland, and J.~M. Martinis.
\newblock {Superconducting quantum circuits at the surface code threshold for
  fault tolerance}.
\newblock \emph{Nature}, 508:\penalty0 500--503, April 2014.
\newblock \doi{10.1038/nature13171}.
\newblock arXiv:1402.4848.

\bibitem[Barker et~al.(2018)Barker, Chen, Roginsky, Vassilev, and
  Davis]{nist-sp-800-56-part1-rev3-2018}
E.~Barker, L.~Chen, A.~Roginsky, A.~Vassilev, and R.~Davis.
\newblock {Recommendation for Pair-Wise Key-Establishment Schemes Using
  Discrete Logarithm Cryptography}.
\newblock Technical Report NIST Special Publication (SP) 800-56A, Rev.~3, NIST,
  April 2018.
\newblock \doi{10.6028/NIST.SP.800-56Ar3}.

\bibitem[Barker et~al.(2019)Barker, Chen, Roginsky, Vassilev, Davis, and
  Simon]{nist-sp-800-56-part2-rev2-2018}
E.~Barker, L.~Chen, A.~Roginsky, A.~Vassilev, R.~Davis, and S.~Simon.
\newblock {Recommendation for Pair-Wise Key Establishment Using Integer
  Factorization Cryptography}.
\newblock Technical Report NIST Special Publication (SP) 800-56B, Rev.~2, NIST,
  March 2019.
\newblock \doi{10.6028/NIST.SP.800-56Br2}.

\bibitem[Beauregard(2003)]{beauregard2002shor}
S.~Beauregard.
\newblock {Circuit for Shor's algorithm using $2n+3$ qubits}.
\newblock \emph{{Quantum Information \& Computation}}, 3\penalty0 (2):\penalty0
  175--185, 2003.
\newblock \doi{10.26421/QIC3.2-8}.
\newblock arXiv:quant-ph/0205095.

\bibitem[Beckman et~al.(1996)Beckman, Chari, Devabhaktuni, and
  Preskill]{beckman1996efficient}
D.~Beckman, A.~N. Chari, S.~Devabhaktuni, and J.~Preskill.
\newblock {Efficient networks for quantum factoring}.
\newblock \emph{{Physical Review A}}, 54\penalty0 (2):\penalty0 1034, 1996.
\newblock \doi{10.1103/PhysRevA.54.1034}.
\newblock arXiv:quant-ph/9602016.

\bibitem[Berry et~al.(2019)Berry, Gidney, Motta, McClean, and
  Babbush]{berry2019qubitization}
D.~W. Berry, C.~Gidney, M.~Motta, J.~R. McClean, and R.~Babbush.
\newblock {Qubitization of Arbitrary Basis Quantum Chemistry Leveraging
  Sparsity and Low Rank Factorization}.
\newblock \emph{{Quantum}}, 3:\penalty0 208, 2019.
\newblock \doi{10.22331/q-2019-12-02-208}.
\newblock arXiv:1902.02134.

\bibitem[{BlueKrypt}(2019)]{keylength2019}
{BlueKrypt}.
\newblock {Cryptographic Key Length Recommendation}.
\newblock \url{https://www.keylength.com}, 2019.
\newblock URL \url{https://www.keylength.com}.
\newblock {Accessed: 2019-03-03}.

\bibitem[Bocharov et~al.(2015)Bocharov, Roetteler, and Svore]{bocharov2015rus}
A.~Bocharov, M.~Roetteler, and K.~M. Svore.
\newblock {Efficient Synthesis of Universal Repeat-Until-Success Quantum
  Circuits}.
\newblock \emph{{Physical Review Letters}}, 114:\penalty0 080502, Feb 2015.
\newblock \doi{10.1103/PhysRevLett.114.080502}.
\newblock arXiv:1404.5320.

\bibitem[Boudot et~al.(2020)Boudot, Gaudry, Guillevic, Heninger, Thomé, and
  Zimmermann]{boudot-rsa829}
F.~Boudot, P.~Gaudry, A.~Guillevic, N.~Heninger, E.~Thomé, and P.~Zimmermann.
\newblock {Comparing the Difficulty of Factorization and Discrete Logarithm: A
  240-Digit Experiment}.
\newblock In \emph{{Advances in Cryptology -- CRYPTO 2020}}, volume 12171 of
  \emph{{Lecture Notes in Computer Science (LNCS)}}, pages 62--91. Springer,
  2020.
\newblock \doi{10.1007/978-3-030-56880-1_3}.

\bibitem[Braithwaite(2016)]{google2016postquantum}
M.~Braithwaite.
\newblock {Experimenting with post-quantum cryptography}.
\newblock
  \url{https://security.googleblog.com/2016/07/experimenting-with-post-quantum.html},
  July 2016.
\newblock URL
  \url{https://security.googleblog.com/2016/07/experimenting-with-post-quantum.html}.

\bibitem[Bravyi and Kitaev(2005)]{bravyi2005distillation}
S.~Bravyi and A.~Kitaev.
\newblock {Universal quantum computation with ideal Clifford gates and noisy
  ancillas}.
\newblock \emph{{Physical Review A}}, 71\penalty0 (2):\penalty0 022316, 2005.
\newblock \doi{10.1103/PhysRevA.71.022316}.
\newblock arXiv:quant-ph/0403025.

\bibitem[Buhler et~al.(1993)Buhler, Lenstra~Jr., and Pomerance]{buhler-nfs}
J.~P. Buhler, H.~W. Lenstra~Jr., and C.~Pomerance.
\newblock {Factoring integers with the number field sieve}.
\newblock In \emph{{The Development of the Number Field Sieve}}, volume 1554 of
  \emph{{Lecture Notes in Mathematics (LNM)}}, pages 50--94. Springer, 1993.
\newblock \doi{10.1007/BFb0091539}.

\bibitem[Campbell et~al.(2019)Campbell, Khurana, and
  Montanaro]{campbell2018constraintsatisfaction}
E.~Campbell, A.~Khurana, and A.~Montanaro.
\newblock {Applying quantum algorithms to constraint satisfaction problems}.
\newblock \emph{{Quantum}}, 3:\penalty0 167, 2019.
\newblock \doi{10.22331/q-2019-07-18-167}.
\newblock arXiv:1810.05582.

\bibitem[Cleve and Watrous(2000)]{cleve2000fast}
R.~Cleve and J.~Watrous.
\newblock {Fast parallel circuits for the quantum Fourier transform}.
\newblock In \emph{{Proceedings 41st Annual Symposium on Foundations of
  Computer Science}}, pages 526--536. IEEE, 2000.
\newblock \doi{10.1109/SFCS.2000.892140}.

\bibitem[Copsey et~al.(2003)Copsey, Oskin, Impens, Metodiev, Cross, Chong,
  Chuang, and Kubiatowicz]{copsey2003toward}
D.~Copsey, M.~Oskin, F.~Impens, T.~Metodiev, A.~Cross, F.~T. Chong, I.~L.
  Chuang, and J.~Kubiatowicz.
\newblock {Toward a scalable, silicon-based quantum computing architecture}.
\newblock \emph{{IEEE Journal of Selected Topics in Quantum Electronics}},
  9\penalty0 (6):\penalty0 1552--1569, 2003.
\newblock \doi{10.1109/JSTQE.2003.820922}.

\bibitem[Cuccaro et~al.(2004)Cuccaro, Draper, Kutin, and
  Moulton]{cuccaro2004adder}
S.~A. Cuccaro, T.~G. Draper, S.~A. Kutin, and D.~P. Moulton.
\newblock {A new quantum ripple-carry addition circuit}.
\newblock \emph{arXiv preprint quant-ph/0410184}, 2004.
\newblock URL \url{https://arxiv.org/abs/quant-ph/0410184}.

\bibitem[Diffie and Hellman(1976)]{diffie-hellman}
W.~Diffie and M.~E. Hellman.
\newblock {New Directions in Cryptography}.
\newblock \emph{{IEEE Transactions on Information Theory}}, IT-22\penalty0
  (6):\penalty0 644--654, 1976.
\newblock \doi{10.1109/TIT.1976.1055638}.

\bibitem[Draper et~al.(2006)Draper, Kutin, Rains, and
  Svore]{draper2004logarithmic}
T.~G. Draper, S.~A. Kutin, E.~M. Rains, and K.~M. Svore.
\newblock {A logarithmic-depth quantum carry-lookahead adder}.
\newblock \emph{{Quantum Information \& Computation}}, 6\penalty0
  (4--5):\penalty0 351--369, 2006.
\newblock \doi{10.26421/QIC6.4-5-4}.
\newblock arXiv:quant-ph/0406142.

\bibitem[Eker{\aa}(2016)]{ekeraa2016modifying}
M.~Eker{\aa}.
\newblock {Modifying Shor's algorithm to compute short discrete logarithms}.
\newblock \emph{Cryptology ePrint Archive, Report 2016/1128}, 2016.
\newblock URL \url{https://eprint.iacr.org/2016/1128}.

\bibitem[Eker{\aa}(2019)]{ekeraa2019revisiting}
M.~Eker{\aa}.
\newblock {Revisiting Shor’s quantum algorithm for computing general discrete
  logarithms}.
\newblock \emph{arXiv preprint arXiv:1905.09084}, 2019.
\newblock URL \url{https://arxiv.org/abs/1905.09084}.

\bibitem[Eker{\aa}(2020)]{ekeraa2017pp}
M.~Eker{\aa}.
\newblock {On post-processing in the quantum algorithm for computing short
  discrete logarithms}.
\newblock \emph{{Designs, Codes and Cryptography}}, 88\penalty0 (11):\penalty0
  2313--2335, 2020.
\newblock \doi{10.1007/s10623-020-00783-2}.
\newblock iacr:2017/1122.

\bibitem[Eker{\aa}(2021)]{ekeraa2018general}
M.~Eker{\aa}.
\newblock {Quantum algorithms for computing general discrete logarithms and
  orders with tradeoffs}.
\newblock \emph{{Journal of Mathematical Cryptology}}, 15\penalty0
  (1):\penalty0 359--407, 2021.
\newblock \doi{10.1515/jmc-2020-0006}.
\newblock (To appear.) iacr:2018/797.

\bibitem[Eker{\aa} and H{\aa}stad(2017)]{ekeraa2017quantum}
M.~Eker{\aa} and J.~H{\aa}stad.
\newblock {Quantum Algorithms for Computing Short Discrete Logarithms and
  Factoring RSA Integers}.
\newblock In \emph{{Post-Quantum Cryptography}}, volume 10346 of \emph{{Lecture
  Notes in Computer Science (LNCS)}}, pages 347--363. Springer, 2017.
\newblock \doi{10.1007/978-3-319-59879-6_20}.

\bibitem[Fowler(2012)]{fowler2012time}
A.~G. Fowler.
\newblock {Time-optimal quantum computation}.
\newblock \emph{arXiv preprint arXiv:1210.4626}, 2012.
\newblock URL \url{https://arxiv.org/abs/1210.4626}.

\bibitem[Fowler and Gidney(2018)]{fowler2018}
A.~G. Fowler and C.~Gidney.
\newblock {Low overhead quantum computation using lattice surgery}.
\newblock \emph{arXiv preprint arXiv:1808.06709}, 2018.
\newblock URL \url{https://arxiv.org/abs/1808.06709}.

\bibitem[Fowler et~al.(2012)Fowler, Mariantoni, Martinis, and
  Cleland]{fowler2012surfacecodereview}
A.~G. Fowler, M.~Mariantoni, J.~M. Martinis, and A.~N. Cleland.
\newblock {Surface codes: Towards practical large-scale quantum computation}.
\newblock \emph{{Physical Review A}}, 86\penalty0 (3):\penalty0 032324, 2012.
\newblock \doi{10.1103/PhysRevA.86.032324}.
\newblock arXiv:1208.0928.

\bibitem[Fowler et~al.(2013)Fowler, Devitt, and Jones]{fowler2013surfaceblock}
A.~G. Fowler, S.~J. Devitt, and C.~Jones.
\newblock {Surface code implementation of block code state distillation}.
\newblock \emph{{Scientific Reports}}, 3:\penalty0 1939, 2013.
\newblock \doi{10.1038/srep01939}.
\newblock arXiv:1301.7107.

\bibitem[Gheorghiu and Mosca(2019)]{gheorghiu2019cryptanalysis}
V.~Gheorghiu and M.~Mosca.
\newblock {Benchmarking the quantum cryptanalysis of symmetric, public-key and
  hash-based cryptographic schemes}.
\newblock \emph{arXiv preprint arXiv:1902.02332}, 2019.
\newblock URL \url{https://arxiv.org/abs/1902.02332}.

\bibitem[Gidney(2017)]{gidney2017factoring}
C.~Gidney.
\newblock {Factoring with $n+2$ clean qubits and $n-1$ dirty qubits}.
\newblock \emph{arXiv preprint arXiv:1706.07884}, 2017.
\newblock URL \url{https://arxiv.org/abs/1706.07884}.

\bibitem[Gidney(2018)]{gidney2018addition}
C.~Gidney.
\newblock {Halving the cost of quantum addition}.
\newblock \emph{{Quantum}}, 2:\penalty0 74, 2018.
\newblock \doi{10.22331/q-2018-06-18-74}.
\newblock arXiv:1709.06648.

\bibitem[Gidney(2019{\natexlab{a}})]{gidney2019approximatepermutation}
C.~Gidney.
\newblock {Approximate encoded permutations and piecewise quantum adders}.
\newblock \emph{arXiv preprint arXiv:1905.08488}, 2019{\natexlab{a}}.
\newblock URL \url{https://arxiv.org/abs/1905.08488}.

\bibitem[Gidney(2019{\natexlab{b}})]{gidney2019karatsuba}
C.~Gidney.
\newblock {Asymptotically Efficient Quantum Karatsuba Multiplication}.
\newblock \emph{arXiv preprint arXiv:1904.07356}, 2019{\natexlab{b}}.
\newblock URL \url{https://arxiv.org/abs/1904.07356}.

\bibitem[Gidney(2019{\natexlab{c}})]{gidney2019windowedarithmetic}
C.~Gidney.
\newblock {Windowed quantum arithmetic}.
\newblock \emph{arXiv preprint arXiv:1905.07682}, 2019{\natexlab{c}}.
\newblock URL \url{https://arxiv.org/abs/1905.07682}.

\bibitem[Gidney and Fowler(2019{\natexlab{a}})]{gidney2018magic}
C.~Gidney and A.~G. Fowler.
\newblock {Efficient magic state factories with a catalyzed
  $|\text{CCZ}\rangle$ to $2|\text{T}\rangle$ transformation}.
\newblock \emph{{Quantum}}, 3:\penalty0 135, 2019{\natexlab{a}}.
\newblock \doi{10.22331/q-2019-04-30-135}.
\newblock arXiv:1812.01238.

\bibitem[Gidney and Fowler(2019{\natexlab{b}})]{gidney2019autoccz}
C.~Gidney and A.~G. Fowler.
\newblock {Flexible layout of surface code computations using AutoCCZ states}.
\newblock \emph{arXiv preprint arXiv:1905.08916}, 2019{\natexlab{b}}.
\newblock URL \url{https://arxiv.org/abs/1905.08916}.

\bibitem[Gillmor(2016)]{rfc-tls}
D.~Gillmor.
\newblock {RFC 7919: Negotiated Finite Field Diffie-Hellman Ephemeral
  Parameters for Transport Layer Security (TLS)}, August 2016.
\newblock \doi{10.17487/RFC7919}.

\bibitem[Gordon(1993)]{gordon}
D.~M. Gordon.
\newblock {Discrete logarithms in GF($p$) using the Number Field Sieve}.
\newblock \emph{{SIAM Journal on Discrete Mathematics}}, 6\penalty0
  (1):\penalty0 124--138, 1993.
\newblock \doi{10.1137/0406010}.

\bibitem[Griffiths and Niu(1996)]{griffiths1996semiclassical}
R.~B. Griffiths and C.-S. Niu.
\newblock {Semiclassical Fourier Transform for Quantum Computation}.
\newblock \emph{{Physical Review Letters}}, 76\penalty0 (17):\penalty0
  3228--3231, April 1996.
\newblock \doi{10.1103/PhysRevLett.76.3228}.
\newblock arXiv:quant-ph/9511007.

\bibitem[Haah and Hastings(2018)]{haah2018codes}
J.~Haah and M.~B. Hastings.
\newblock {Codes and Protocols for Distilling $T$, controlled-$S$, and Toffoli
  Gates}.
\newblock \emph{{Quantum}}, 2:\penalty0 71, 2018.
\newblock \doi{10.22331/q-2018-06-07-71}.
\newblock arXiv:1709.02832.

\bibitem[H{\"a}ner et~al.(2017)H{\"a}ner, Roetteler, and
  Svore]{haner2016factoring}
T.~H{\"a}ner, M.~Roetteler, and K.~M. Svore.
\newblock {Factoring using $2n+2$ qubits with Toffoli based modular
  multiplication}.
\newblock \emph{{Quantum Information \& Computation}}, 17\penalty0
  (7--8):\penalty0 673--684, 2017.
\newblock \doi{10.26421/QIC17.7-8-7}.
\newblock arXiv:1611.07995.

\bibitem[Hastings and Geller(2015)]{hastings2014dislocations}
M.~B. Hastings and A.~Geller.
\newblock {Reduced Space-Time and Time Costs Using Dislocation Codes and
  Arbitrary Ancillas}.
\newblock \emph{{Quantum Information \& Computation}}, 15\penalty0
  (11--12):\penalty0 962–986, 2015.
\newblock \doi{10.26421/QIC15.11-12-6}.
\newblock arXiv:1408.3379.

\bibitem[Horsman et~al.(2012)Horsman, Fowler, Devitt, and
  Van~Meter]{horsman2012latticesurgery}
C.~Horsman, A.~G. Fowler, S.~Devitt, and R.~Van~Meter.
\newblock {Surface code quantum computing by lattice surgery}.
\newblock \emph{{New Journal of Physics}}, 14\penalty0 (12):\penalty0 123011,
  2012.
\newblock \doi{10.1088/1367-2630/14/12/123011}.
\newblock arXiv:1111.4022.

\bibitem[Jiang et~al.(2010)Jiang, Taylor, S{\o}rensen, and
  Lukin]{jiang2010scalable}
L.~Jiang, J.~M. Taylor, A.~S. S{\o}rensen, and M.~D. Lukin.
\newblock {Scalable quantum networks based on few-qubit registers}.
\newblock \emph{{International Journal of Quantum Information}}, 8\penalty0
  (01n02):\penalty0 93--104, 2010.
\newblock \doi{10.1142/S0219749910006058}.

\bibitem[Jones et~al.(2012)Jones, Van~Meter, Fowler, McMahon, Kim, Ladd, and
  Yamamoto]{jones2012layered}
N.~C. Jones, R.~Van~Meter, A.~G. Fowler, P.~L. McMahon, J.~Kim, T.~D. Ladd, and
  Y.~Yamamoto.
\newblock {Layered Architecture for Quantum Computing}.
\newblock \emph{{Physical Review X}}, 2\penalty0 (3):\penalty0 031007, 2012.
\newblock \doi{10.1103/PhysRevX.2.031007}.
\newblock arXiv:1010.5022.

\bibitem[Karatsuba and Ofman(1962)]{karatsuba1962multiplication}
A.~A. Karatsuba and Y.~P. Ofman.
\newblock {Multiplication of many-digital numbers by automatic computers}.
\newblock \emph{{Doklady Akademii Nauk SSSR}}, 145\penalty0 (2):\penalty0
  293--294, 1962.
\newblock URL \url{http://mi.mathnet.ru/eng/dan/v145/i2/p293}.

\bibitem[Kim et~al.(2014)Kim, Daly, Kim, Fallin, Lee, Lee, Wilkerson, Lai, and
  Mutlu]{Kim2014}
Y.~Kim, R.~Daly, J.~Kim, C.~Fallin, J.~H. Lee, D.~Lee, C.~Wilkerson, K.~Lai,
  and O.~Mutlu.
\newblock {Flipping bits in memory without accessing them: An experimental
  study of DRAM disturbance errors}.
\newblock In \emph{{2014 ACM/IEEE 41st International Symposium on Computer
  Architecture (ISCA)}}, pages 361--372. IEEE, 2014.
\newblock \doi{10.1109/ISCA.2014.6853210}.

\bibitem[Kivinen and Kojo(2003)]{rfc-ike}
T.~Kivinen and M.~Kojo.
\newblock {RFC 3526: More Modular Exponentiation (MODP) Diffie-Hellman groups
  for Internet Key Exchange (IKE)}, May 2003.
\newblock \doi{10.17487/RFC3526}.

\bibitem[Kleinjung et~al.(2010)Kleinjung, Aoki, Franke, Lenstra, Thomé, Bos,
  Gaudry, Kruppa, Montgomery, Osvik, Herman, Timofeev, and
  Zimmermann]{kleinjung-rsa768}
T.~Kleinjung, K.~Aoki, J.~Franke, A.~K. Lenstra, E.~Thomé, J.~W. Bos,
  P.~Gaudry, A.~Kruppa, P.~L. Montgomery, D.~A. Osvik, t.~R. Herman,
  A.~Timofeev, and P.~Zimmermann.
\newblock {Factorization of a 768-Bit RSA Modulus}.
\newblock In \emph{{Advances in Cryptology -- CRYPTO 2010}}, volume 6223 of
  \emph{{Lecture Notes in Computer Science (LNCS)}}, pages 333--350. Springer,
  2010.
\newblock \doi{10.1007/978-3-642-14623-7_18}.

\bibitem[Kutin(2006)]{kutin2006shor}
S.~A. Kutin.
\newblock {Shor's algorithm on a nearest-neighbor machine}.
\newblock \emph{arXiv preprint quant-ph/0609001}, 2006.
\newblock URL \url{https://arxiv.org/abs/quant-ph/0609001}.

\bibitem[Lenstra(2004)]{lenstra-model-2004}
A.~K. Lenstra.
\newblock {Key Lengths}.
\newblock In \emph{{The Handbook of Information Security}}, chapter~6. 2004.
\newblock URL
  \url{https://infoscience.epfl.ch/record/164539/files/NPDF-32.pdf}.

\bibitem[Lenstra and Lenstra~Jr.(1993)]{nfs-book}
A.~K. Lenstra and H.~W. Lenstra~Jr., editors.
\newblock \emph{{The Development of the Number Field Sieve}}, volume 1554 of
  \emph{{Lecture Notes in Mathematics (LNM)}}.
\newblock Springer, 1993.
\newblock \doi{10.1007/BFb0091534}.

\bibitem[Lenstra and Verheul(2001)]{lenstra-verheul-model-2001}
A.~K. Lenstra and E.~R. Verheul.
\newblock {Selecting Cryptographic Key Sizes}.
\newblock \emph{{Journal of Cryptology}}, 14:\penalty0 225--293, 2001.
\newblock \doi{10.1007/s00145-001-0009-4}.

\bibitem[Lenstra et~al.(1990)Lenstra, Lenstra~Jr., Manasse, and
  Pollard]{lenstra-nfs}
A.~K. Lenstra, H.~W. Lenstra~Jr., M.~S. Manasse, and J.~M. Pollard.
\newblock {The number field sieve}.
\newblock In \emph{{Proceedings of the Twenty-Second Annual ACM Symposium on
  Theory of Computing}}, pages 564--572. ACM, 1990.
\newblock \doi{10.1145/100216.100295}.

\bibitem[Litinski(2019)]{litinski2019magic}
D.~Litinski.
\newblock {Magic State Distillation: Not as Costly as You Think}.
\newblock \emph{{Quantum}}, 3:\penalty0 205, 2019.
\newblock \doi{10.22331/q-2019-12-02-205}.
\newblock arXiv:1905.06903.

\bibitem[Mosca(2018)]{mosca2018cybersecurity}
M.~Mosca.
\newblock {Cybersecurity in an Era with Quantum Computers: Will We Be Ready?}
\newblock \emph{{IEEE Security \& Privacy}}, 16\penalty0 (5):\penalty0 38--41,
  2018.
\newblock \doi{10.1109/MSP.2018.3761723}.
\newblock iacr:2015/1075.

\bibitem[Mosca and Ekert(1999)]{mosca1999recycle}
M.~Mosca and A.~Ekert.
\newblock {The Hidden Subgroup Problem and Eigenvalue Estimation on a Quantum
  Computer}.
\newblock In \emph{{Quantum Computing and Quantum Communications}}, volume 1509
  of \emph{{Lecture Notes in Computer Science (LNCS)}}, pages 174--188.
  Springer, 1999.
\newblock \doi{10.1007/3-540-49208-9_15}.

\bibitem[NIST(2013)]{fips-186-4}
NIST.
\newblock {Digital Signature Standard (DSS)}.
\newblock Technical Report Federal Information Processing Standards
  Publications (FIPS PUBS) 186-4, July 2013.
\newblock \doi{10.6028/NIST.FIPS.186-4}.

\bibitem[NIST and CCCS(2019)]{fips-140-2-IG}
NIST and CCCS.
\newblock {Implementation Guidance for FIPS 140-2 and the Cryptographic Module
  Validation Program}.
\newblock Technical report, May 2019.
\newblock URL
  \url{https://csrc.nist.gov/csrc/media/projects/cryptographic-module-validation-program/documents/fips140-2/fips1402ig.pdf}.
\newblock {Accessed: 2019-05-10, Document Revision: 2019-05-07}.

\bibitem[O'Gorman and Campbell(2017)]{ogorman2017factories}
J.~O'Gorman and E.~T. Campbell.
\newblock {Quantum computation with realistic magic-state factories}.
\newblock \emph{{Physical Review A}}, 95\penalty0 (3):\penalty0 032338, 2017.
\newblock \doi{10.1103/PhysRevA.95.032338}.
\newblock arXiv:1605.07197.

\bibitem[Oi et~al.(2006)Oi, Devitt, and Hollenberg]{oi2006scalable}
D.~K.~L. Oi, S.~J. Devitt, and L.~C.~L. Hollenberg.
\newblock {Scalable error correction in distributed ion trap computers}.
\newblock \emph{{Physical Review A}}, 74\penalty0 (5):\penalty0 052313, 2006.
\newblock \doi{10.1103/PhysRevA.74.052313}.
\newblock arXiv:quant-ph/0606226.

\bibitem[{OpenSSL Software Foundation}(2018)]{open-ssl-source2018}
{OpenSSL Software Foundation}.
\newblock {OpenSSL source code: Line 32 of \texttt{apps/dhparam.c}}.
\newblock
  \url{https://github.com/openssl/openssl/blob/07f434441e7ea385f975e8df8caa03e62222ca61/apps/dhparam.c\#L32},
  2018.
\newblock URL
  \url{https://github.com/openssl/openssl/blob/07f434441e7ea385f975e8df8caa03e62222ca61/apps/dhparam.c\#L32}.
\newblock {Accessed: 2018-12-11}.

\bibitem[Oskin et~al.(2002)Oskin, Chong, and Chuang]{oskin2002practical}
M.~Oskin, F.~T. Chong, and I.~L. Chuang.
\newblock {A practical architecture for reliable quantum computers}.
\newblock \emph{{Computer}}, 35\penalty0 (1):\penalty0 79--87, 2002.
\newblock \doi{10.1109/2.976922}.

\bibitem[Parent et~al.(2018)Parent, Roetteler, and Mosca]{parent2017karatsuba}
A.~Parent, M.~Roetteler, and M.~Mosca.
\newblock {Improved reversible and quantum circuits for Karatsuba-based integer
  multiplication}.
\newblock In \emph{{12th Conference on the Theory of Quantum Computation,
  Communication and Cryptography (TQC 2017)}}, volume~73 of \emph{{Leibniz
  International Proceedings in Informatics (LIPIcs)}}, pages 7:1--7:15.
  {Schloss Dagstuhl -- Leibniz-Zentrum für Informatik}, 2018.
\newblock \doi{10.4230/LIPIcs.TQC.2017.7}.
\newblock arXiv:1706.03419.

\bibitem[Parker and Plenio(2000)]{parker2000recycle}
S.~Parker and M.~B. Plenio.
\newblock {Efficient Factorization with a Single Pure Qubit and $\log N$ Mixed
  Qubits}.
\newblock \emph{{Physical Review Letters}}, 85\penalty0 (14):\penalty0 3049,
  October 2000.
\newblock \doi{10.1103/PhysRevLett.85.3049}.
\newblock arXiv:quant-ph/0001066.

\bibitem[Pavlidis and Gizopoulos(2014)]{pavlidis2012fast}
A.~Pavlidis and D.~Gizopoulos.
\newblock {Fast quantum modular exponentiation architecture for Shor's
  factorization algorithm}.
\newblock \emph{{Quantum Information \& Computation}}, 14\penalty0
  (7--8):\penalty0 649--682, 2014.
\newblock \doi{10.26421/QIC14.7-8-8}.
\newblock arXiv:1207.0511.

\bibitem[Pohlig and Hellman(1978)]{pohlig-hellman}
S.~C. Pohlig and M.~E. Hellman.
\newblock {An Improved Algorithm for Computing Logarithms over GF($p$) and Its
  Cryptographic Significance}.
\newblock \emph{{IEEE Transactions on Information Theory}}, IT-24\penalty0
  (1):\penalty0 106--110, 1978.
\newblock \doi{10.1109/TIT.1978.1055817}.

\bibitem[Pollard(1978)]{pollard-rho-lambda}
J.~M. Pollard.
\newblock {Monte Carlo Methods for Index Computation (mod $p$)}.
\newblock \emph{{Mathematics of Computation}}, 32\penalty0 (143):\penalty0
  918--924, 1978.
\newblock \doi{10.1090/s0025-5718-1978-0491431-9}.

\bibitem[Pollard(1993{\natexlab{a}})]{pollard-cubic}
J.~M. Pollard.
\newblock {Factoring with cubic integers}.
\newblock In \emph{{The Development of the Number Field Sieve}}, volume 1554 of
  \emph{{Lecture Notes in Mathematics (LNM)}}, pages 4--10. Springer,
  1993{\natexlab{a}}.
\newblock \doi{10.1007/BFb0091536}.

\bibitem[Pollard(1993{\natexlab{b}})]{pollard-nfs-sieving}
J.~M. Pollard.
\newblock {The lattice sieve}.
\newblock In \emph{{The Development of the Number Field Sieve}}, volume 1554 of
  \emph{{Lecture Notes in Mathematics (LNM)}}, pages 43--49. Springer,
  1993{\natexlab{b}}.
\newblock \doi{10.1007/BFb0091538}.

\bibitem[Pomerance(1996)]{pomerance1996atale}
C.~Pomerance.
\newblock {A Tale of Two Sieves}.
\newblock \emph{{Notices of the AMS}}, 43\penalty0 (12):\penalty0 1473--1485,
  1996.
\newblock URL \url{https://www.ams.org/notices/199612/pomerance.pdf}.

\bibitem[Rivest et~al.(1978)Rivest, Shamir, and Adleman]{rsa}
R.~L. Rivest, A.~Shamir, and L.~Adleman.
\newblock {A Method for Obtaining Digital Signatures and Public-Key
  Cryptosystems}.
\newblock \emph{{Communications of the ACM}}, 21\penalty0 (2):\penalty0
  120--126, 1978.
\newblock \doi{10.1145/359340.359342}.

\bibitem[Roetteler et~al.(2017)Roetteler, Naehrig, Svore, and
  Lauter]{roetteler2017quantum}
M.~Roetteler, M.~Naehrig, K.~M. Svore, and K.~Lauter.
\newblock {Quantum Resource Estimates for Computing Elliptic Curve Discrete
  Logarithms}.
\newblock In \emph{{Advances in Cryptology -- ASIACRYPT 2017}}, volume 10625 of
  \emph{{Lecture Notes in Computer Science (LNCS)}}, pages 241--270. Springer,
  2017.
\newblock \doi{10.1007/978-3-319-70697-9_9}.

\bibitem[Schirokauer(1992)]{schirokauer-thesis}
O.~Schirokauer.
\newblock \emph{{On pro-finite groups and on discrete logarithms}}.
\newblock PhD thesis, University of California, Berkeley, May 1992.

\bibitem[Schirokauer(1993)]{schirokauer}
O.~Schirokauer.
\newblock {Discrete Logarithms and Local Units}.
\newblock \emph{{Philosophical Transactions of the Royal Society of London A}},
  345\penalty0 (1676):\penalty0 409--423, 1993.
\newblock \doi{10.1098/rsta.1993.0139}.

\bibitem[Sch{\"o}nhage and Strassen(1971)]{schonhage1971multiply}
A.~Sch{\"o}nhage and V.~Strassen.
\newblock {Schnelle Multiplikation großer Zahlen}.
\newblock \emph{{Computing}}, 7\penalty0 (3--4):\penalty0 281--292, 1971.
\newblock \doi{10.1007/BF02242355}.

\bibitem[Schroeder et~al.(2009)Schroeder, Pinheiro, and
  Weber]{schroeder2009dram}
B.~Schroeder, E.~Pinheiro, and W.-D. Weber.
\newblock {DRAM Errors in the Wild: A Large-Scale Field Study}.
\newblock \emph{{SIGMETRICS Performance Evaluation Review}}, 37\penalty0
  (1):\penalty0 193--204, 2009.
\newblock \doi{10.1145/1555349.1555372}.

\bibitem[Shor(1994)]{shor1994}
P.~W. Shor.
\newblock {Algorithms for Quantum Computation: Discrete Logarithms and
  Factoring}.
\newblock In \emph{{Proceedings 35th Annual Symposium on Foundations of
  Computer Science}}, pages 124--134. IEEE, 1994.
\newblock \doi{10.1109/SFCS.1994.365700}.

\bibitem[{The GnuPG Project}(2018)]{gpg-faq-key-size2018}
{The GnuPG Project}.
\newblock {GnuPG Frequently Asked Questions: Why does GnuPG default to 2048 bit
  RSA-2048?}
\newblock \url{https://www.gnupg.org/faq/gnupg-faq.html\#default_rsa2048},
  2018.
\newblock URL \url{https://www.gnupg.org/faq/gnupg-faq.html\#default_rsa2048}.
\newblock {Accessed: 2018-12-11}.

\bibitem[{The OpenSSH Project}(2018)]{ssh-keygen-man-page2018}
{The OpenSSH Project}.
\newblock {Linux Documentation: Man Page for \texttt{ssh-keygen(1)}}.
\newblock \url{https://linux.die.net/man/1/ssh-keygen}, 2018.
\newblock URL \url{https://linux.die.net/man/1/ssh-keygen}.
\newblock {Accessed: 2018-12-11}.

\bibitem[Van~Meter(2019)]{vanquantumcomputerarchitecture}
R.~Van~Meter.
\newblock {A \#QuantumComputerArchitecture Tweetstorm}, 2019.
\newblock \doi{10.5281/zenodo.3496597}.

\bibitem[Van~Meter and Itoh(2005)]{van2005fastexponentiation}
R.~Van~Meter and K.~M. Itoh.
\newblock {Fast quantum modular exponentiation}.
\newblock \emph{{Physical Review A}}, 71\penalty0 (5):\penalty0 052320, 2005.
\newblock \doi{10.1103/PhysRevA.71.052320}.
\newblock arXiv:quant-ph/0408006.

\bibitem[Van~Meter et~al.(2008)Van~Meter, Munro, Nemoto, and
  Itoh]{meter2008arithmetic}
R.~Van~Meter, W.~J. Munro, K.~Nemoto, and K.~M. Itoh.
\newblock {Arithmetic on a distributed-memory quantum multicomputer}.
\newblock \emph{{ACM Journal on Emerging Technologies in Computing Systems
  (JETC)}}, 3\penalty0 (4):\penalty0 1--23, 2008.
\newblock \doi{10.1145/1324177.1324179}.

\bibitem[Van~Meter et~al.(2010)Van~Meter, Ladd, Fowler, and
  Yamamoto]{van2010distributed}
R.~Van~Meter, T.~D. Ladd, A.~G. Fowler, and Y.~Yamamoto.
\newblock {Distributed quantum computation architecture using semiconductor
  nanophotonics}.
\newblock \emph{{International Journal of Quantum Information}}, 8\penalty0
  (01n02):\penalty0 295--323, 2010.
\newblock \doi{10.1142/S0219749910006435}.
\newblock arXiv:0906.2686.

\bibitem[{van}~Oorschot and Wiener(1996)]{oorschot}
P.~C. {van}~Oorschot and M.~J. Wiener.
\newblock {On Diffie-Hellman Key Agreement with Short Exponents}.
\newblock In \emph{{Advances in Cryptology -- EUROCRYPT '96}}, volume 1070 of
  \emph{{Lecture Notes in Computer Science (LNCS)}}, pages 332--343. Springer,
  1996.
\newblock \doi{10.1007/3-540-68339-9_29}.

\bibitem[Vedral et~al.(1996)Vedral, Barenco, and Ekert]{vedral1996arithmetic}
V.~Vedral, A.~Barenco, and A.~Ekert.
\newblock {Quantum networks for elementary arithmetic operations}.
\newblock \emph{{Physical Review A}}, 54\penalty0 (1):\penalty0 147--153, 1996.
\newblock \doi{10.1103/PhysRevA.54.147}.
\newblock arXiv:quant-ph/9511018.

\bibitem[Whitney et~al.(2009)Whitney, Isailovic, Patel, and
  Kubiatowicz]{whitney2009fault}
M.~G. Whitney, N.~Isailovic, Y.~Patel, and J.~Kubiatowicz.
\newblock {A fault tolerant, area efficient architecture for Shor's factoring
  algorithm}.
\newblock In \emph{{Proceedings of the 36th Annual International Symposium on
  Computer Architecture}}, pages 383--394. ACM, 2009.
\newblock \doi{10.1145/1555754.1555802}.

\bibitem[{Wikipedia}(2018)]{Timeline}
{Wikipedia}.
\newblock {Timeline of quantum computing}.
\newblock \url{https://en.wikipedia.org/wiki/Timeline_of_quantum_computing},
  2018.
\newblock URL
  \url{https://en.wikipedia.org/wiki/Timeline_of_quantum_computing}.
\newblock {Accessed: 2018-12-18}.

\bibitem[Zalka(1998)]{zalka1998fast}
C.~Zalka.
\newblock {Fast versions of Shor's quantum factoring algorithm}.
\newblock \emph{arXiv preprint quant-ph/9806084}, 1998.
\newblock URL \url{https://arxiv.org/abs/quant-ph/9806084}.

\bibitem[Zalka(2006)]{zalka2006pure}
C.~Zalka.
\newblock {Shor's algorithm with fewer (pure) qubits}.
\newblock \emph{arXiv preprint quant-ph/0601097}, 2006.
\newblock URL \url{https://arxiv.org/abs/quant-ph/0601097}.

\end{thebibliography}
\appendix

\section{Notes on \texorpdfstring{\tbl{comparison}}{Table I}}
\label{app:table-details}

\subsection{Columns}

\begin{itemize}
    \item \textbf{Abstract Qubits}:
        The number of logical qubits used in the abstract circuit model.
        Ignores qubits used for distillation and routing.
    \item \textbf{Measurement Depth}:
        The length of the longest chain of dependent measurements, which determines the reaction limited runtime of an algorithm.
        These numbers are not adjusted to account for the chance of retrying.
    \item \textbf{Toffoli+T/2}:
        Number of magic states required by the algorithm.
        The ``/2" adjustment is intended to account for the fact that T states take less volume to distill than Toffoli states.
        These numbers are not adjusted to account for the chance of retrying.
    \item \textbf{Min volume}:
        Expected spacetime cost of the computation, including retries.
        The papers included in the table span decades, and a range of assumptions about the architecture of quantum computers, and generally do not provide spacetime volumes.
        To assign volumes to these papers, we plugged their asymptotic formulas into various possible realizations of that algorithm (serial, parallel, and intermediate) using ancillary file ``fill-in-table.py".
        Each volume entry is the minimum volume achieved by the different possible realizations.
        We assumed all papers but our own had a negligible retry chance.
\end{itemize}

\subsection{Entries}

Some entries in the table are directly from a paper, others had to be inferred by hand, and others were filled in using the output of the ancillary file ``fill-in-table.py".
Here is where each entry in the table came from:

\begin{itemize}
    \item Vedral et al. 1996 \cite{vedral1996arithmetic}:
    \begin{itemize}
        \item Toffoli+T/2 Count:
            $80n^3$ derived from figures at end of the paper.
        \item Abstract Qubits:
            Paper says $7n+1$ (end of section IV).
            Verified using figures at end of paper.
            Paper mentions this can be improved to $4n + 3$, but the described method would radically increase the Toffoli count.
        \item Measurement Depth:
            Equal to the Toffoli count.
            The circuit construction is serial.
        \item Numbers at specific $n$: from ``fill-in-table.py".
    \end{itemize}
    \item Zalka 1998 (basic) \cite{zalka1998fast}:
    \begin{itemize}
        \item Toffoli+T/2 Count:
            Paper says $12n^3$ (title of section 1.4.1).
        \item Abstract Qubits:
            Paper says $3n$ (title of section 1.4.1).
        \item Measurement Depth:
            Equal to the Toffoli count.
            The circuit construction is serial.
        \item Numbers at specific $n$: from ``fill-in-table.py".
    \end{itemize}
    \item Zalka 1998 (log add) \cite{zalka1998fast}:
    \begin{itemize}
        \item Toffoli+T/2 Count:
            Paper says $52n^3$ (section 5, bottom of page 18).
        \item Abstract Qubits:
            Paper says $5n$ (section 5, bottom of page 18).
        \item Measurement Depth:
            Paper says $600n^2$ (section 5, top of page 19).
        \item Numbers at specific $n$: from ``fill-in-table.py".
    \end{itemize}
    \item Zalka 1998 (fft mult) \cite{zalka1998fast}:
    \begin{itemize}
        \item Toffoli+T/2 Count:
            Paper says $2^{17}n^2$ (section 5, middle of page 19).
        \item Abstract Qubits:
            Paper says $96n$ (section 5, middle of page 19).
        \item Measurement Depth:
            Paper says $2^{17}n^{1.2}$ (section 5, middle of page 19).
        \item Numbers at specific $n$: from ``fill-in-table.py".
    \end{itemize}
    \item Beauregard 2002 \cite{beauregard2002shor}:
    \begin{itemize}
        \item Toffoli+T/2 Count: Derived from figures in the paper. Our estimate has an additional factor of $\lg n$ to account for the need to approximate arbitrary phase rotations using T states.
        \item Abstract Qubits: Stated in the title of the paper.
        \item Measurement Depth: Derived from figures in the paper. Our estimate has an additional factor of $\lg n$ to account for the need to approximate arbitrary phase rotations using T states.
        \item Numbers at specific $n$: from ``fill-in-table.py".
    \end{itemize}
    \item Fowler et al. 2012 \cite{fowler2012surfacecodereview}:
    \begin{itemize}
        \item Toffoli+T/2 Count: From table I in the paper.
        \item Abstract Qubits: The paper erroneously claims $2n$ in table I due to overlooking a necessary workspace register.
        We corrected this to $3n + O(1)$.
        \item Measurement Depth: From table I in the paper.
        \item Numbers at specific $n$: from ``fill-in-table.py".
    \end{itemize}
    \item H\"{a}ner et al. 2016 \cite{haner2016factoring}:
    \begin{itemize}
        \item Toffoli+T/2 Count:
            From table 2 of \cite{roetteler2017quantum}.
        \item Abstract Qubits:
            From paper's title.
        \item Measurement Depth:
            Derived $52n^3$ from the circuit diagrams included in the paper.
        \item Numbers at specific $n$: from ``fill-in-table.py".
    \end{itemize}
    \item (ours) 2019:
        Asymptotic bounds explained in \sec{construction}.
        Concrete numbers at specific sizes come from ancillary file ``estimate\_costs.py".
        The Toffoli count does not account for the chance of retrying (because it is mostly insensitive to changes that lower this chance), but the volume does account for it (i.e. it is the expected total volume to factor, not the per-run volume).
    \item Roetteler et al. 2017 \cite{roetteler2017quantum}:
        \begin{itemize}
            \item Toffoli+T/2 Count (asymptotic):
                From the paper's abstract.
            \item Abstract Qubits:
                From the paper's abstract.
            \item Measurement Depth:
                Assumed same as Toffoli count.
            \item Toffoli+T/2 Count (at $n=224$):
                From Table 2.
            \item Numbers at specific $n$: from ``fill-in-table.py".
        \end{itemize}
        Note that an $n$ bit prime order elliptic curve group provides approximately $n/2$~bits of classical security (see Appendix~D to \cite{nist-sp-800-56-part1-rev3-2018}), since the best classical attacks are cycle-finding attacks,
        whilst RSA-1024, RSA-2048 and RSA-3072 provide approximately $80$, $112$ and $128$~bits of classical security, respectively, according to the model used by NIST (see Appendix D to~\cite{nist-sp-800-56-part2-rev2-2018}).
        There are several other models~\cite{keylength2019, lenstra-model-2004, lenstra-verheul-model-2001}.
        The table compares problem instances that provide the same level of classical security according to the NIST model.
\end{itemize}

\section{Notes on \texorpdfstring{\tbl{historical-comparison}}{Table II}}
\label{app:historical-table-details}

\subsection{Columns}

\begin{itemize}
    \item \textbf{Physical gate error rate}:
        The probability that executing a physical gate will introduce Pauli errors onto targeted qubits.
    \item \textbf{Cycle time}:
        The amount of time it takes to measure all of the surface code's stabilizers once.
    \item \textbf{Reaction time}:
        The amount of time it takes the classical control system to trigger a logical measurement, collect and error-correct the result, and decide on which measurement basis to use for the next set of measurements.
    \item \textbf{Physical connectivity}:
        Which physical qubits can interact with each other.
        Planar means only adjacent qubits on a planar grid can interact (typical of superconducting qubits).
        Arbitrary means qubits can interact with other qubits as needed by the construction (typical of ion traps).
    \item \textbf{Distillation strategy}:
        The dominant kind of magic states being distilled, and the number or type of factory being used to distill them.
    \item \textbf{Execution strategy}:
        Notes on how the computation progresses at a low level.
        Each construction spends most of its time performing additions, so this column ended up describing addition strategies.
        The suffixes ``carry lookahead", ``ripple carry", and ``carry runways" describe the type of adder being used.
        The prefixes ``distillation limited" and ``reaction limited" describe what is preventing the adder from running faster.
        Distillation limited means that the adder is bottlenecked on magic states being produced.
        Reaction limited means that the adder is bottlenecked on the control system resolving long chains of dependent measurement bases.
    \item \textbf{Physical qubits}:
        The number of physical qubits used by the computation, using the original historical assumptions.
    \item \textbf{Expected runtime}:
        The average amount of time before the computation has completed successfully, using the original historical assumptions.
    \item \textbf{Expected volume}:
        The average number of physical qubit-rounds before the computation has completed successfully, using the original historical assumptions.
\end{itemize}

\subsection{Entries}

\begin{itemize}
    \item Van Meter et al. 2009 \cite{van2010distributed}:
        Numbers are derived from Table 2 of \cite{van2010distributed}.
        The paper does not use reaction limited computation, so the reaction time is not included.
    \item Jones et al. 2009 \cite{jones2012layered}:
        Numbers are derived from Figure 15, Table II, and Table VII of \cite{jones2012layered}.
        The paper does not use reaction limited computation, so the reaction time is not included.
    \item Fowler et al. 2012 \cite{fowler2012surfacecodereview}:
        Numbers are from the background section of \cite{fowler2012surfacecodereview}.
        This paper's expected volume is directly comparable to ours, despite the different reaction time assumption, because their volume estimate is dominated by distillation and changing the reaction time does not affect distillation volume.
    \item O'Gorman et al. (2017) \cite{ogorman2017factories}:
        Numbers are derived from table I of \cite{ogorman2017factories}.
        The physical qubit count of $2.18 \cdot 10^8$ from their table only includes distillation; we increased it by $3 \cdot 2048 \cdot (2 \cdot 32^2) \approx 0.13 \cdot 10^8$ to account for three $n$-logical-qubit data registers.
    \item Gheorghiu et al. (2019) \cite{gheorghiu2019cryptanalysis}
        The physical gate error rate, qubit count, and time are stated in the caption of fig 45 (``B. RSA-2048") of the paper.
        The surface code cycle time is stated to be 200 nanoseconds in Section B, soon after equation 3.
        The T factory count was inferred by dividing the stated T count of $2.4 \cdot 10^{12}$ by the production rate of the T factory from~\cite{fowler2018}, which is approximately 25kHz for the given surface code cycle time.
        The paper does not state a reaction time, but the T states are being used to perform chains of dependent Toffolis.
        This implies the reaction time must be faster than the number factories times the per-factory production rate divided by 4 (the number of Ts needed to perform a Toffoli).
        So a 0.2 microsecond reaction time would not be sufficient, but a 0.1 microsecond reaction time would be, and so we state the reaction time as 0.1 microseconds.
        \\
    \item Ours (2019):
        The parallel entry is the implementation described in \sec{construction} and produced by ancillary file ``estimates\_costs.py".
        The single threaded and serial distillation entries use the same basic architecture, but with fewer factories and without piecewise additions or double-speed lookups.
\end{itemize}
\end{document}